\documentclass[%
preprint,
 amsmath,amssymb,
 aps,
]{revtex4-2}

\usepackage{graphicx}
\usepackage{dcolumn}
\usepackage{bm}
\usepackage[mathlines]{lineno}

\usepackage{xcolor}
\usepackage{cancel}
\usepackage{soul}
\usepackage{booktabs}
\usepackage{multirow}

\definecolor{ForestGreen}{RGB}{34,139,34}

\newcommand{\luciano}[1]{\textcolor{red}{#1}}

\begin{document}



\title{
{New results on proton-induced reactions on Vanadium for $^{\mbox{\rm 47}}$Sc production and the impact of level densities on theoretical cross sections}}

\author{F. Barbaro$^{1,2}$, L. Canton$^{1}$, M.P. Carante$^{2,3}$, A. Colombi$^{2,3}$, L. De Dominicis$^{4,5}$, A. Fontana$^{3}$, F. Haddad$^{6,7}$, L. Mou$^{5}$, G. Pupillo$^{5}$}

\affiliation{$^{1}$INFN Sezione di Padova, via F. Marzolo, 8, I-35131 Padova, Italy \\ $^{2}$Universit\`a di Pavia, Dipartimento di Fisica, via A. Bassi 6, I-27100 Pavia, Italy \\ $^{3}$INFN Sezione di Pavia, via A. Bassi 6, I-27100 Pavia, Italy \\ $^{4}$Universit\`a di Padova, Dipartimento di Fisica e Astronomia, via F. Marzolo, 8, I-35131 Padova, Italy \\ $^{5}$ INFN Laboratori Nazionali di Legnaro, Viale dell'Universit\`a, Legnaro, Padova, I-35020 , Italy\\
$^{6}$ Subatech, CNRS/IN2P3, IMT Atlantique, Université de Nantes, CS 20722, 44307, Nantes Cedex, France\\
$^{7}$ GIP ARRONAX, Rue Arronax 1, 44800, Saint-Herblain, France}


\date{\today}

\begin{abstract}

New data for the $^{\mbox{\rm nat}}$V(p,x) reactions have been measured in the range {26-70} MeV, with production of the nuclides  $^{47}$Sc, $^{43}$Sc, $^{44m}$Sc, $^{44g}$Sc, $^{46}$Sc, $^{48}$Sc, $^{42}$K, $^{43}$K, $^{48}$V, $^{48}$Cr, $^{49}$Cr, and
$^{51}$Cr.
The focus is on the production of $^{47}$Sc, 
a $\beta^-$-emitter suitable for innovative radiotheranostic applications in nuclear medicine. The measured cross sections for this radionuclide
and its contaminants are compared with the theoretical excitation functions calculated with the TALYS code. 
In view of novel radiopharmaceutical applications, 
it is essential to accurately describe these cross-sections for the evaluation of yields, purities,  and dose releases. Hence, we optimize the level-density parameters of the microscopic models in the TALYS code to obtain the best possible descriptions of the new data.
We consider different irradiation conditions to estimate the production yields from the cross sections determined in this work.

\end{abstract}

\maketitle


\section{Introduction}

The theranostic radionuclide $^{47}$Sc has gained the attention of the scientific community for its
favourable decay characteristics ($T_{1/2}$=3.35 days, $E_{\beta^-}$=162 keV, $E_{\gamma}$=159 keV, I=68.4\%) \cite{NNDC} 
that makes it one of the most attractive radionuclides for nuclear
medicine applications~\cite{jalilian2020iaea}. In addition to its long half-life, suitable to follow the slow biodistribution of large
molecules such as monoclonal antibodies, $^{47}$Sc can be used for SPECT imaging, thanks to its
$\gamma$-emission, and to treat small-size tumours, thanks to its high intensity low-range $\beta^-$ radiation. The stable
coordination of Sc element with the chelating agent DOTA paves the way to new
$^{47}$Sc-radiopharmaceuticals, while the possibility to pair $^{47}$Sc with its
positron-emitter counterparts, $^{43}$Sc and $^{44}$Sc, permits also to perform low dose PET
studies before, during, and after therapy \cite{CRP, Muller2018,Loveless2019}. Despite some promising preclinical results, the use of
$^{47}$Sc-labelled radiopharmaceuticals in nuclear medicine is curtailed by the lack of
$^{47}$Sc availability in sufficiently high yield and medically acceptable purity \cite{Qaim2020}. For this
reason, all possible $^{47}$Sc production routes are investigated worldwide, considering both accelerators
(cyclotrons and LINACs) and nuclear reactors. The PASTA project (acronym for Production with Accelerator of Sc-47 for
Theranostic Applications), aimed at measuring the most
promising nuclear reactions to produce $^{47}$Sc considering proton-beams \cite{pasta2019, cimento2018}.
The project was developed in the framework of the LARAMED program at the INFN-Legnaro National Laboratories (LNL), where a 70 MeV proton cyclotron was
installed and the infrastructure for nuclear cross section measurements and small-scale production of medical
radionuclides for preclinical applications is under completion \cite{laramed2019}.

The fundamental role of nuclear physics in the production of radionuclides for medical applications is highlighted by
several authors \cite{Qaim2020,knapp2016,IAEAreport,CRP}. One of the major goals is to minimize the co-production of
contaminant isotopes during the irradiation since they cannot be separated with a radiochemical process. In the case of
$^{47}$Sc, the main contaminant is $^{46}$Sc due to its long half-life (83.79 d) but the
contribution to the final dose imparted to the patient has to be carefully analysed considering the {specificity of each radiopharmaceutical} and all the Scandium radionuclides co-produced \cite{denardo2021}. In this context nuclear reaction models are of great aid to describe the cross sections, particularly for radionuclides or contaminants that are difficult to be measured. For
example, in {the case} of $^{47}$Sc production, attention has to be paid to the co-production of $^{45}$Sc (stable) and $^{49}$Sc, which emits low-intensity $\gamma$ lines. Both nuclides could hardly be observed, but nevertheless they require an estimation since they
affect the final isotopic purity and specific activity of the Sc-labelled radiopharmaceutical. 

This work is a compendium of the nuclear cross sections measured in the PASTA project by using proton beams and $^{nat}$V targets, namely $^{47}$Sc and Sc-isotopes ($^{43}$Sc, $^{44m}$Sc, $^{44g}$Sc, $^{46}$Sc, $^{48}$Sc), $^{42}$K and $^{43}$K, $^{48}$V and $^{48}$Cr, $^{49}$Cr and $^{51}$Cr. The experimental values regarding the production of these twelve radionuclides are compared with the literature. In particular, the case of $^{47}$Sc was studied since 1956 by Heininger \cite{Heininger1956}, whose data cover the 60-240 MeV energy range. However, those results show the maximum experimentally observed deviation from the average literature and large uncertainties \cite{exfor}. The values of the $^{nat}$V(p,x)$^{47}$Sc cross section obtained by Hontzeas and Yaffe in 1963 are very scattered and seem to not properly reproduce the trend of the reaction \cite{Hontzeas1963}; the peak, measured at 42 MeV, is overestimated and the more recent measurements agree on the evaporation peak at lower energies ($E_p$ $<$ 40 MeV). The group by Michel et al. measured the proton-induced reactions on $^{nat}$V targets in two experimental campaigns: the first, covering the 18.5-45 MeV range was published in 1979 \cite{michel1979}, the second, covering the 50-200 MeV range was published in 1985 \cite{michel1985}. {The $^{47}$Sc production cross section from these measurements is also in agreement with the data at low energies (E $<$ 30MeV) by Levkovski \cite{Levkovskij1991}.} The more recent data, provided by Ditroi et al. in 2016, cover the 36-64 MeV energy window and were compared with results obtained by using EMPIRE and TALYS nuclear codes \cite{ditroi2016}. Similarly, the experimental values obtained within the PASTA project for the 26.5-70 MeV energy range are here compared with results obtained by using state-of-art nuclear models {that will be discussed futher on.} The results will allow to accurately estimate the production yields of $^{47}$Sc and contaminants and to design an irradiation experiment for the production of $^{47}$Sc with the highest isotopic and radionuclidic purities. A preliminary assessment of the optimal conditions was presented in \cite{denardo2021} in which the energy interval 19-30 MeV was identified as the most promising by performing a {parametric} extrapolation at low energy of the available data: {in this analysis we follow a different approach by considering a microscopic theoretical model and, by introducing the tuning of the level densities, we get an excellent reproduction of the data in the region (mass and energy) of interest. Thus, without any parametric-fit extrapolation of data at low energies, we obtain an optimized  theoretical prediction of the yields, as well as of the contamination by stable isotopes, such as $^{45}$Sc, which typically are not measurable by $\gamma$-spectroscopy. The advantage is that a complete theoretical model has built-in the accurate constraints (centrifugal and Coulomb repulsion, and mass-threshold conditions, etc.) that are needed for a correct threshold behaviour, while a simple fit extrapolation may be not so accurate.}


{Nuclear reaction codes incorporate a statistical evaporation module, generally based upon the Hauser-Feshbach theory~\cite{canton-fontana-2020}. Amongst the various aspects that enter in that formalism, a crucial role is played by the Nuclear Levels Density (NLD), characterizing the excitation structure of the compound nuclei. Up to the last century, the description of NLD was based mainly upon phenomenological models that could lead to analytical forms quite flexible but depending largely on the adjustment of parameters to the available data. Many phenomenological NLD models have been developed along these lines, starting from the basic Fermi-Gas model, to the composite Gilbert-Cameron model~\cite{gilbert-cameron-1965}, or to Generalized-superfluid model\cite{ignatyuk-1993}.}

{It was only starting from the years 2000 that the microscopic descriptions became sufficiently accurate in the global NLD description, for an extremely large set of nuclides (about 8000/8500), to compete with the phenomenological models for practical applications. One of the first works was developed in Ref.\cite{demetriou-goriely-2001}, where effective Skyrme-type nucleon-nucleon interactions have been employed in the context of mean-field theories. The method used the Hartree-Fock plus Bardeen–Cooper–Schrieffer model (HF-BCS) to describe the many-body structure and to incorporate the pairing force. NLD was then calculated using a microscopic statistical method, as described also in Ref.\cite{minato-2011}.}

{Further advancements in microscopic NLD calculations were achieved in Refs.~\cite{hilaire2006,goriely2008}, by resorting to a more consistent (axially symmetric) Hartree-Fock-Bogoliubov (HFB) scheme with Skyrme interaction, with the NLD constructed using a shell-model combinatorial approach~\cite{hillman-grover-1969}.
This lead to quite a few improvements, such as the determination of the parity dependence of the NLD, together with the inclusion of the collective aspects of both rotational and vibrational character in the theoretical description.}

More recently, a new study has been completed~\cite{hilaire2012} with improvements in the microscopic description of the collective effects, both vibrational and rotational, based on the use of a new Gogny-type nucleon-nucleon effective interaction in the HFB approach. Also, the introduction of a temperature-dependent generalization allowed to take into account the evolution of the nuclear deformation with increasing excitation energy, an important aspect that has been disregarded in previous treatments.

{
Other important aspects of the theoretical description of nuclear reactions are the optical potential model used to describe the interaction and the preequilibrium model. Talys provides state-of-art implementations for both and we have made calculations with all the available options. We have obtained excellent results with the microscopic model JLM (Jeukenne-Lejeune-Mahaux) for the optical potential, as explained in the following, and with the exciton model with numerical transition rates (default option) for preequilibrium. In all the calculations, the optical potential and preequilibrium models are used ``as provided", while for the NLD we have obtained a significant improvement with the scaling procedure described in Section \ref{NLDscaling}. 
}

This nuclear reaction study is devoted to a specific practical application of nuclear medicine: the evaluation of
the production route of the theranostic $^{47}$Sc radionuclide,  along with its main contaminants, using the cyclotron-based $^{\mbox{\rm nat}}$V(p,x) channel. This application-oriented problem demands a model description of cross sections as close as possible to the experimental data, since many quantities (yields, radionuclidic purities, etc.) are needed with high accuracy for the multidisciplinary evaluation of the production route. {Indeed, a precise determination of the relevant quantities is important for the prediction of the amount of radio-pharmaceutical compound that can be produced in view of pre-clinical and clinical trials, 
for the evaluation of the dosimetric impact to the patient's organs, and for the appropriate radiochemistry separation techniques.}

{These practical aspects imply that a good reproduction of the relevant cross sections is much more important than the overall theoretical consistency of models. For this reason we have adjusted the NLD parameters for an optimal reproduction of the cross sections, even if it increases the discrepancy between observed and calculated nuclear-level cumulatives}.

 This work presents the final and complete data set obtained by the PASTA experiment.
{In Sect.II, we illustrate the experimental techniques and procedures employed for the acquisition of the new data. However, for details of the apparatus and methods, we refer also to previous publications \cite{pasta2019,papersc43}  where partial and preliminary data of the experiment have been published. In Sect.III, we illustrate the main considerations upon the models implied by the nuclear reaction code TALYS, and how we have used the NLD parameters to optimize the reproduction of the measured cross-sections. The experimental/theoretical results are presented in Sect.IV, and in Sect.V we discuss their implications, including an evaluation of $^{47}$Sc yield obtainable with a standard irradiation-condition set-up. Conclusive remarks are given in Sect.VI.}

\section{Experimental setup and measurements}

\begin{figure}[!htb]
\includegraphics[scale=0.6]{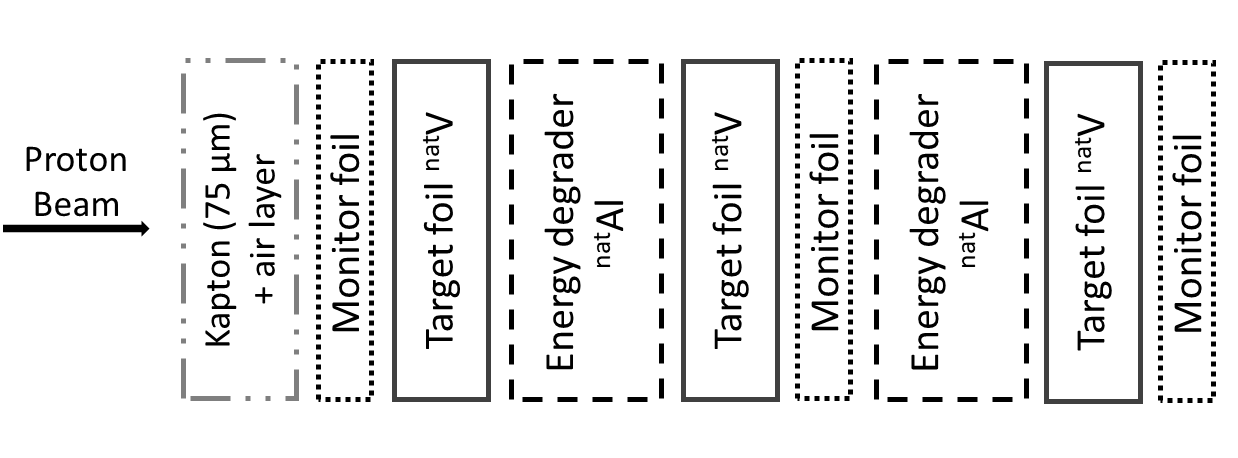}
\caption{Stacked-foils structure.}
\label{fig:1}
\end{figure}

Since the beam-line devoted to cross section measurements for medical radionuclides production is not ready yet at
INFN-LNL, the experiments of the PASTA project were performed at the ARRONAX facility \cite{haddad2008}. Six
irradiation runs were carried out using the low current (100 nA) proton beam with tunable
energy (34-70 MeV), the target-holder and collimator described in a previous work \cite{papersc43}. Stacked-foils targets,
composed by a set of thin metallic $^{\mbox{\rm nat}}$V foils (purity {\textgreater} 99.8\%, thickness=20 {\textmu}m) interchanged with monitor foils and $^{\mbox{\rm nat}}$Al energy degraders, were used in the
experiments \cite{pasta2019}. The $^{\mbox{\rm nat}}$Ni(p,x)$^{57}$Ni and the $^{27}$Al(p,x)$^{24}$Na
cross sections, recommended by the IAEA \cite{IAEAmonitor}, were considered as monitor reactions respectively for energies
lower and higher than 40 MeV. $^{\mbox{\rm nat}}$Ni (purity {\textgreater} 99.95\%, thickness=10 {\textmu}m) or
$^{\mbox{\rm nat}}$Al (purity {\textgreater} 99.0\%, thickness=10 {\textmu}m) foils were thus added in the stacked-structure after each
$^{\mbox{\rm nat}}$V target foil (see Fig. \ref{fig:1}). 
The target was
placed under normal atmosphere downstream the end of the beam line, kept under vacuum and closed with a 75
$\mu $m thick Kapton foil. The distance from the target holder to
the Kapton foil, ranging from 10 cm to 15 cm, was accurately measured for each run. The proton beam energy entering
each layer of the stacked-foils target was calculated using the SRIM code \cite{SRIM}. In order to calculate the energy
losses in the Kapton foil, in the air and across each target foil, the input parameters were the proton beam energy
extracted from the cyclotron and each layer's thickness.
{At the end, at each foil corresponds one energy value, calculated as the mean energy of the input and output energies in the specific layer.}
The uncertainty of the proton beam energy was obtained
considering the uncertainty of the beam energy extracted from the cyclotron (about 500 keV) and by calculating with the 
SRIM code the energy straggling through each layer of the stacked-target.

All samples were measured with the same high-purity germanium (HPGe) detector (model Canberra GC1020, energy resolution FWHM 1.0 keV at 122 keV, relative efficiency 10\%), previously calibrated with
$^{152}$Eu point-like solid source. The sample-detector distance was fixed at 19 cm to keep
the dead time during measurements below 10\%. At least five acquisitions were carried out for each sample, in order to
better follow the decay of the radionuclides of interest. The first measurement of each
$^{\mbox{\rm nat}}$V foil, typically 15 minutes long, started
about 2 hours after the End of Bombardment (EOB); the following acquisitions, usually 3 hours long, were performed once
per day starting from the day after the EOB (up to 5 days). An additional acquisition of each
$^{\mbox{\rm nat}}$V foil was also carried out about 2 months
after the EOB, to further measure the activity of the long-lived
$^{46}$Sc. Table \ref{tab:1} reports the decay
characteristics of all radionuclides of interest, including the reference isotopes produced in the monitor foils, as
extracted from the NuDat 2.8 database \cite{NNDC}.

\begin{table}[!hbt]
{\scriptsize
\renewcommand{\arraystretch}{0.8}
\begin{tabular}{cccccccc}
\toprule
&&\multicolumn{2}{c} {$\gamma$-emission} & \multicolumn{4}{c} {$\beta$-emission} \\ 
\cmidrule(l){3-4} \cmidrule(l){5-8}
Radionuclide & $\tau_{1/2}$ & Energy& Intensity & Mean $\beta^{+}$  & Total $\beta^{+}$  & Mean $\beta^{-}$  & Total $\beta^{-}$  \\
 &   &  (keV) & (\%) &  energy (keV) &  intensity (\%) &  energy (keV) & intensity (\%) \\ 
\midrule
$^{47}$Sc & 3.3492 d & 159.381 & 68.3 & &  & 162.0 & 100.0  \\
$^{43}$Sc & 3.891 h & 372.9 & 22.5 &  476 & 88.1 &  &  \\
$^{44m}$Sc & 58.61 h & 271.241 & 86.74 &  &  &  & \\
 & & 1157.002 & 1.2 & & & & \\
$^{44g}$Sc & 3.97 h & 1157.02 & 99.9 & 632.0 & 94.27 & &   \\
$^{46}$Sc & 83.79 d & 1120.545 & 99.9870 & & & 111.8 & 100.0000  \\
 & & 889.277 & 99.9840 & & & & \\
$^{48}$Sc & 43.67 h & 1312.120 & 100.1 &  & & 220.4 & 100.0 \\ 
 &  & 983.526 & 100.1 & & & &  \\
 & & 1037.522 & 97.6 & & & & \\
$^{42}$K & 12.355 h & 1524.6 & 18.08 & & & 1430.5 & 100.00 \\
$^{43}$K & 22.3 h & 372.760 & 86.80 & & & 317 & 100.0 \\
 & & 617.490 & 79.2 &  &  & & \\
$^{48}$V & 15.9735 d & 983.525 & 99.98 & 290 & 49.9 &  &  \\
 & & 1312.106 & 98.2 & &  & & \\
 & & 944.130 & 7.870 & &  & & \\
$^{48}$Cr & 21.56 h & 308.24 & 100 & 96 & 1.6 &  &  \\
$^{49}$Cr & 42.3 m & 152.928 & 30.3 & 640 & 93 &  &  \\
$^{51}$Cr & 27.704 d & 320.0824 & 9.910 & & & & \\
$^{57}$Ni & 35.60 h & 1377.63 & 81.7 & 354 & 43.6 & &  \\
 & & 127.164 & 16.7 & & & & \\
$^{24}$Na & 14.997 h & 1368.626 & 99.9936 & & & 554.6 & 99.995 \\
\bottomrule
\end{tabular}
}
\caption{Main decay characteristics of the twelve radionuclides of interest and of the two reference isotopes $^{57}$Ni and $^{24}$Na \cite{NNDC}. }
\label{tab:1}
\end{table}

Data analysis, including uncertainty calculations, was performed following the procedure
described in \cite{otuka2017}. {For the case of
$^{44g}$Sc, the activity was obtained by using the
formulation presented in \cite{otuka2017},} that takes 
into account the
$^{44m}$Sc decay correction and the 1157 keV $\gamma
$-line interference as shown in Tab. \ref{tab:1}. The final cross section value related to each target foil was calculated as the
weighted average of all single values associated to each counting; the uncertainty of the monitor cross section was
added at the end of this calculation.

\section{Theoretical calculations}

The theoretical analysis of cross sections for the production of medical radionuclides is important, since a good predictive capability is essential to efficiently guide the experimental activities. Conversely,  new measurements are the key ingredient to validate the 
theoretical models and to optimize the models free parameters. The topic has been thoroughly explored in the past and an exhaustive literature 
exists on the theory of nuclear reactions, ranging from phenomenological models that provide analytical expressions of important observables 
to modern microscopic approaches that aim to a more accurate description of the reaction mechanisms. 

Modern nuclear reaction codes are today the best tools to perform theoretical calculations in this field: in this work we have used the TALYS simulation package \cite{TALYS} which is actively developed and maintained. The aim of the authors is to provide a complete and high quality set of models \cite{utopia} for the different reaction mechanisms that are relevant in the energy range of medical interest, namely nuclear scattering, compound nucleus formation and preequilibrium emission. 


Recently, a great effort was made to implement different microscopic models, both for the optical potential and for the nuclear level densities.
For the optical potential model, we have selected the so-called JLM (Jeukenne-Lejeune-Mahaux) semi-microscopic nucleon-nucleus
 spherical optical model potential as described in  \cite{bauge2001}. This model relies on the 
folding  of a  nucleon-nucleon  effective  interaction in nuclear matter. 
It is based upon the Br\"{u}ckner-Hartree-Fock (BHF) theory to describe the mass operator in infinite nuclear matter, and then applied to finite nuclei 
via the improved local density approximation.
{
To take into account the preequilibrium emission we have used the default option, which is based on the exciton model with transitions rates evaluated numerically.
}
Regarding nuclear levels density, we focus in particular on the microscopic approaches denoted as Hartree-Fock-Models (HFM) which includes both HF-BCS and HFB descriptions, able to provide detailed information about nuclear 
energy levels, as well as masses, spin, parity, and levels density. 
The last parameter has been studied starting from the seminal work of Bethe \cite{bethe} in the context of the compound nucleus theory and in many other phenomenological 
approaches, but from the physical point of view the microscopic approach provides a modern and flexible description of the level densities. 
Indicating with $N(E)$ the cumulative number of levels 
with energy smaller than E, the nuclear level density is defined as \cite{zele}:
\begin{equation}
\rho(E) = \frac{dN(E)}{dE} \ .    
\end{equation}
This quantity can be inferred experimentally from different sources, such as spectroscopic studies, neutron resonances, evaporation spectra and Ericson fluctuations studies \cite{grimes}. Levels density has been calculated in the HFM frameworks and the comparison is usually done by inspecting the cumulative distributions $N(E)$ \cite{ripl}. 

\subsection{Microscopic models in the TALYS code}

Amongst the various options for the choice of the Optical Model in TALYS,  the JLM microscopic optical model, derived according to the work reported 
in \cite{jeukenne1977a,jeukenne1977b}, 
was then adjusted phenomenologically  \cite{bauge2001} to improve the agreement of calculated 
finite-nuclei cross-sections with a large set of experimental data.

Moreover TALYS provides a variety of HFM based microscopic frameworks for the levels density with 3 different tabulated options:
\textit{ldmodel 4} (Microscopic level densities (Skyrme force) from Goriely’s tables {\cite{demetriou-goriely-2001}}), \textit{ldmodel 5} (Microscopic level densities (Skyrme force) from Hilaire’s combinatorial tables \cite{goriely2008}) and \textit{ldmodel 6} ( Microscopic level densities (temperature dependent HFB, Gogny force) from Hilaire’s combinatorial tables \cite{hilaire2012}).

The three models are based on the HFM approach and represent somehow the time evolution of the TALYS code with respect to the microscopic level density
 models, as suggested by the referenced papers. None of them is perfect, but they complement each other being able to describe different aspects of reaction 
mechanisms, possibly on the entire nuclide chart. Each option provides different tabulated level densities, but the code allows to overrule them to improve the 
agreement with data within a restricted Z and A range. In this work we have found good results with the option \textit{ldmodel 4} and with an optimized set of parameters 
that were applied to the calculations by a scaling procedure. 
{The preequilibrium model is fixed in all calculations with the option \textit{preeqmode 2}, except for the $^{51}$Cr cross section where we have sought for an improvement with a more recent addition as discussed in the following.}

\subsection{Nuclear levels density scaling procedure}
\label{NLDscaling}

The approach used in TALYS offers a certain level of flexibility since, in the HFM description, the level densities are tabulated at different excitation energies for both parities and different values of spins: the values in these tables 
can be re-scaled by  simple multiplication to  fit experimental data. This technique has been successfully adopted in \cite{Koning2008} to find a global best 
fit on the entire nuclide chart.

In general the original HFM microscopic level densities have not been adjusted to experimental data and the tabulated values can be rescaled by using two parameters 
($c$ and $p$) according to the transformation: 

\begin{equation}
\rho(E,J,\pi) = exp(c \sqrt{E-p}) \rho_{HFM}(E-p,J,\pi)
\label{eq:1}
\end{equation}

 The parameter $p$ represents a  “pairing shift” and allows to obtain the level density from the table at a different energy. Parameter $c$ acts to the 
level density in similar way  to  that  of  the  parameter $a$ of  phenomenological  models and provides an overall normalisation of the level density. Adjusting $p$ 
and $c$ together provides  flexibility in fitting at both low and high energies, and this implies that both low-energy discrete levels and average experimental 
resonance separations can be described within the same procedure.
This procedure allows to modify the original HFM level density and to obtain different ``customized" tables, as shown in the TALYS manual and on the RIPL library \cite{ripl}.

\subsubsection{Grid search and optimization procedure}

The first step is a grid search on the $c$ and $p$ parameters in the allowed range [-10,10] \cite{TALYSman}, but we have restricted our search in a narrower interval [-2,4] as suggested by Fig.39 in Ref.\cite{ripl}. The grid search is multidimensional, since  it concerns the $c$ and $p$ parameters of all the nuclides involved.
A preliminary solution is found by a first qualitative agreement between theory and measurements, obtained by visual inspection of the cross sections for which new data are available. 
With this trial and error approach, we define the initial values of the $c$ and $p$ parameters for the next step, characterized by the optimization procedure. 

The second step is  to refine the solution by evaluating a global chi-square of all the involved cross-sections: a Minuit-based chi-square minimization procedure \cite{minuit} has been implemented 
by using the $c$ and $p$ values as free parameters, with initial values identified by the previously described grid search. 
The minimization algorithm used is the Minuit recommended one, which is based on the combination of gradient and simplex optimization.
The level density distributions and cumulatives 
associated to the theoretical cross sections that correspond to the minimum chi-square are analyzed and compared with the default HFB and the TALYS proposed ones.

The goal of this work is to achieve the most accurate parametrization of the selected TALYS microscopic model for the cross sections of interest.
Following these steps we have found an optimized set of values for the $c$ and $p$ parameters to fit our experimental cross sections and we have analyzed the corresponding cumulative distributions of levels. 
We report these optimized values, found for the subset of analyzed isotopes, in Tab. \ref{tab:2}.  For a few of the nuclides reported in the table, it was necessary to vary only one parameter ($c$ or $p$), maintaining the other unchanged with respect to the value defined by the Goriely tables adopted in \textit{ldmodel 4}. 
{The solution that we have found in Tab. \ref{tab:2} is not necessarily unique: we cannot exclude that another set of parameters (related for example to a different optical potential, preequilibrium or nuclear level density model) could reproduce the data with a similar accuracy. This would still be in the spirit of this work, \textit{i.e.} to improve the theoretical description of the cross sections of medical radioisotopes and contaminants for radiopharmaceutical applications.}

\begin{table}[!htb]
\begin{tabular}{c|c|c}
Isotope &    $c$ &   $p$ \\ \hline
$^{42}$K & 1.27443 & - \\
$^{43}$K & 1.09559 & 0.140468 \\
$^{44}$Sc & 1.97876 & 1.11855 \\
$^{46}$Sc & 0.339592 & -1.17731 \\
$^{48}$Sc & - & 0.449064 \\
$^{48}$V & - & -0.25841 \\
$^{48}$Cr & 0.234421 & - \\
\end{tabular}
\caption{Optimized $c$ and $p$ parameters used in this work.}
\label{tab:2}
\end{table}

\section{Results}

\begin{figure}[!htb]
\includegraphics[scale=1.0]{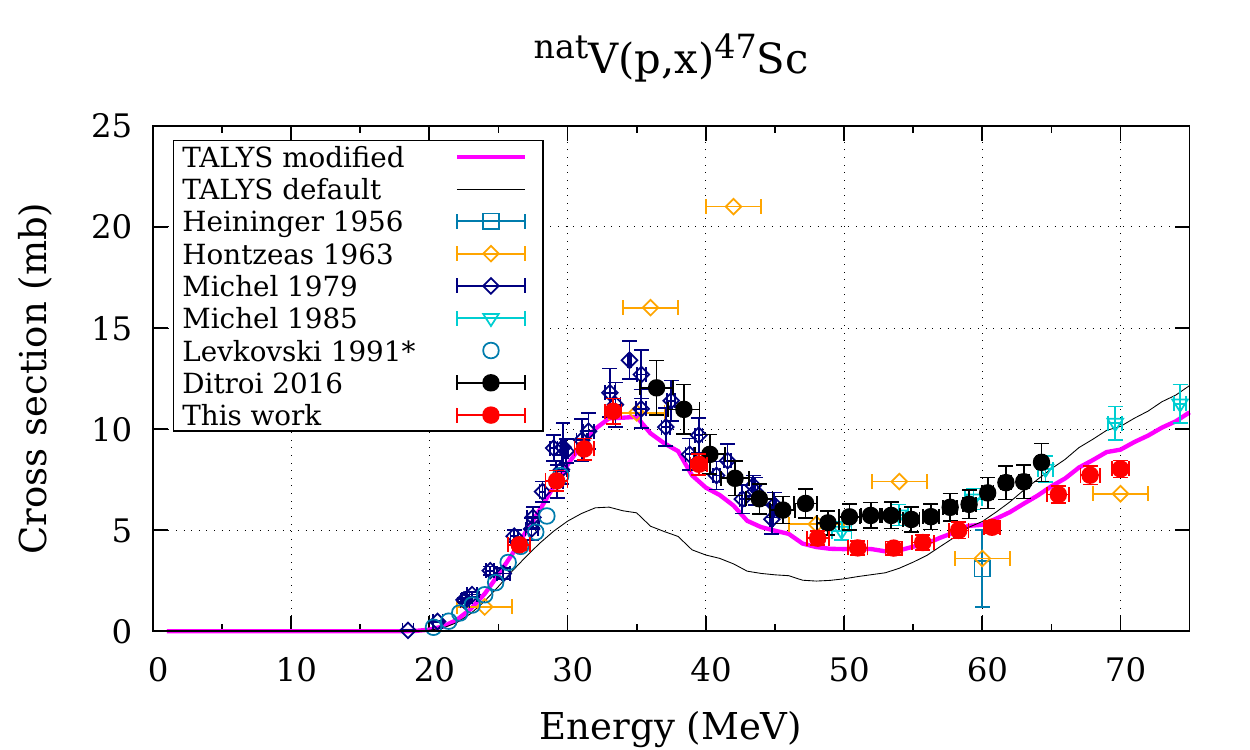}
\caption{\label{fig:sc47xs}$^{47}$Sc cross section.}
\end{figure}

\begin{figure}[!htb]
\includegraphics[scale=1.0]{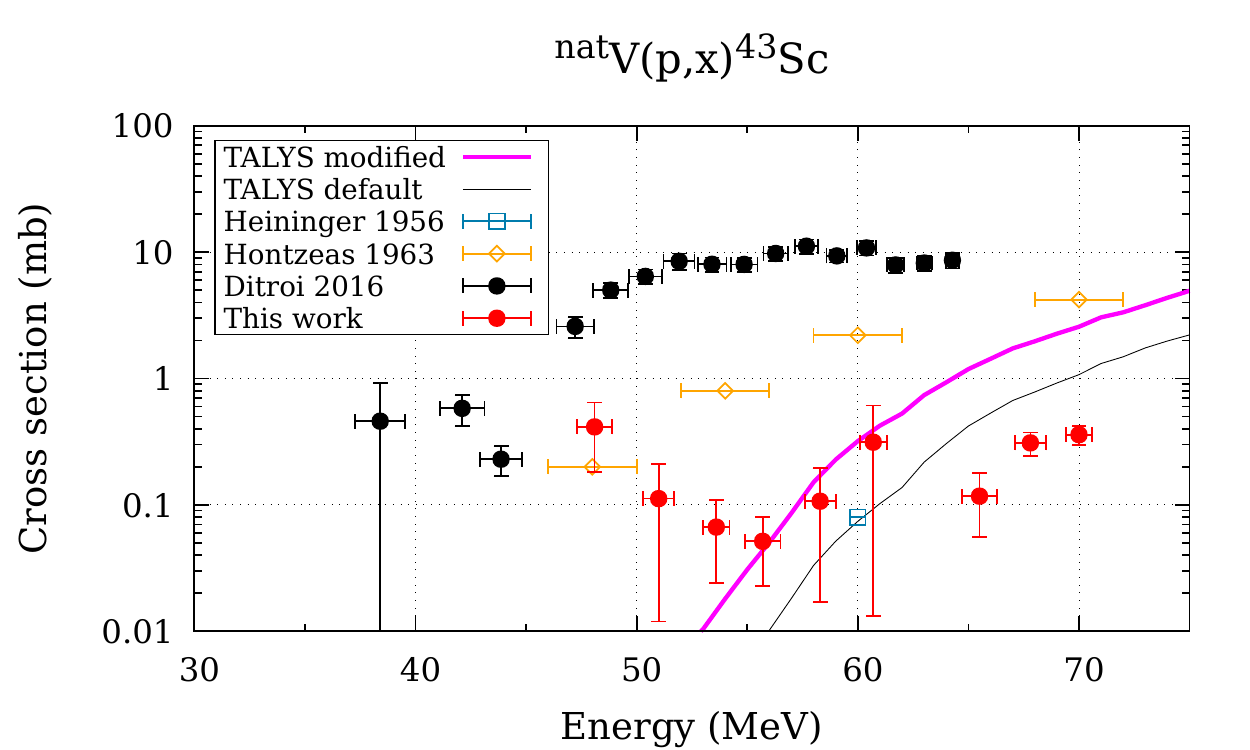}
\caption{\label{fig:sc43xs}$^{43}$Sc cross section.}
\end{figure}

\begin{figure}[!htb]
\includegraphics[scale=1.0]{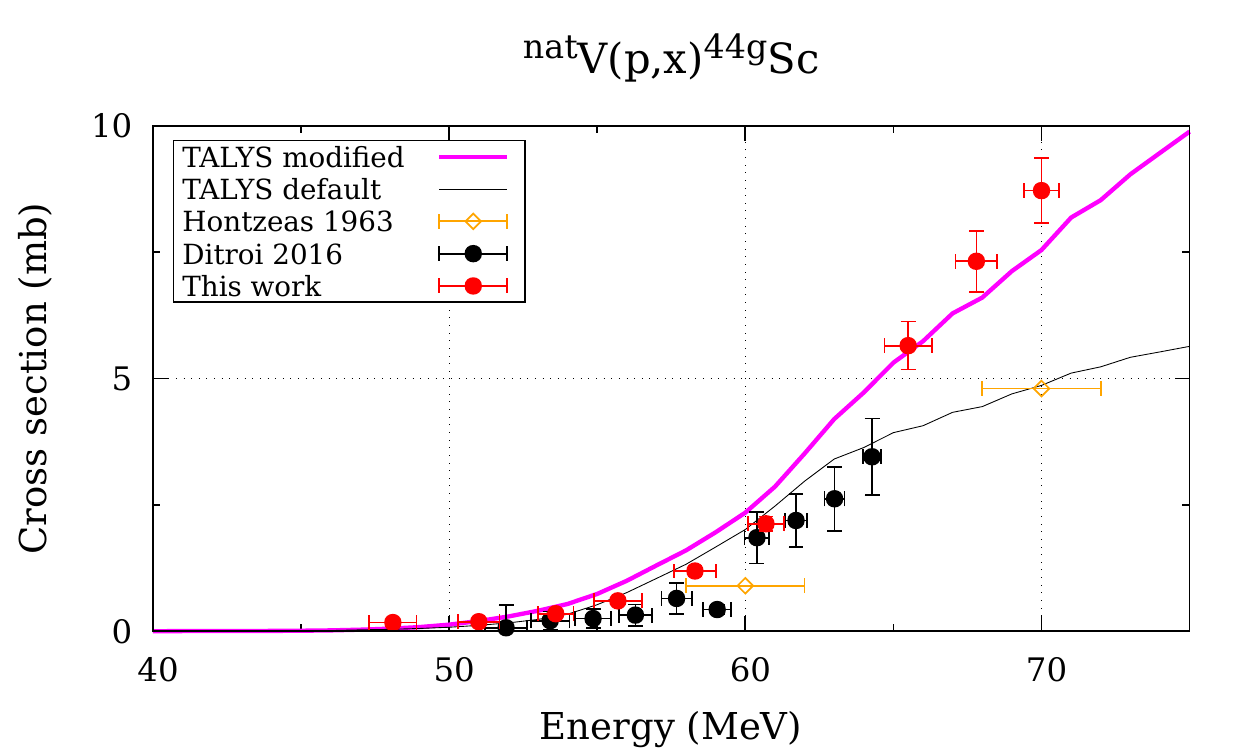}
\caption{\label{fig:sc44gxs}$^{44g}$Sc cross section.}
\end{figure}

\begin{figure}[!htb]
\includegraphics[scale=1.0]{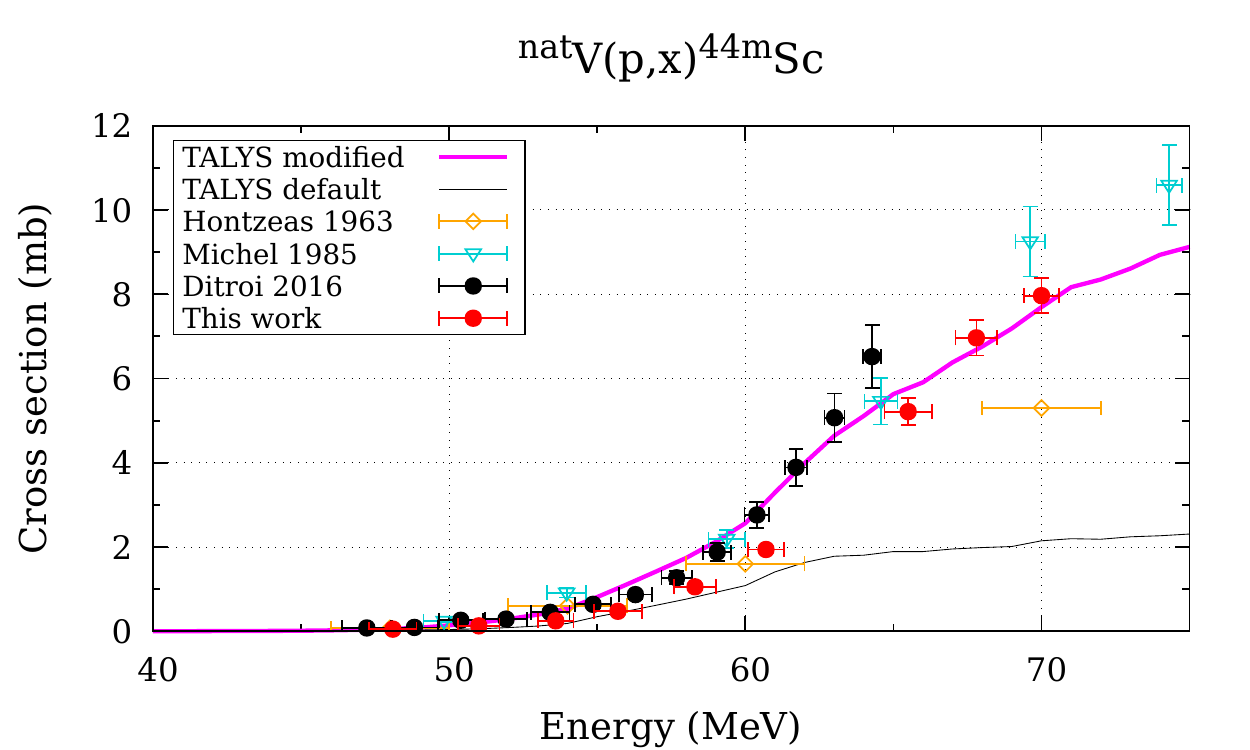}
\caption{\label{fig:sc44mxs}$^{44m}$Sc cross section.}
\end{figure}

\begin{figure}[!htb]
\includegraphics[scale=1.0]{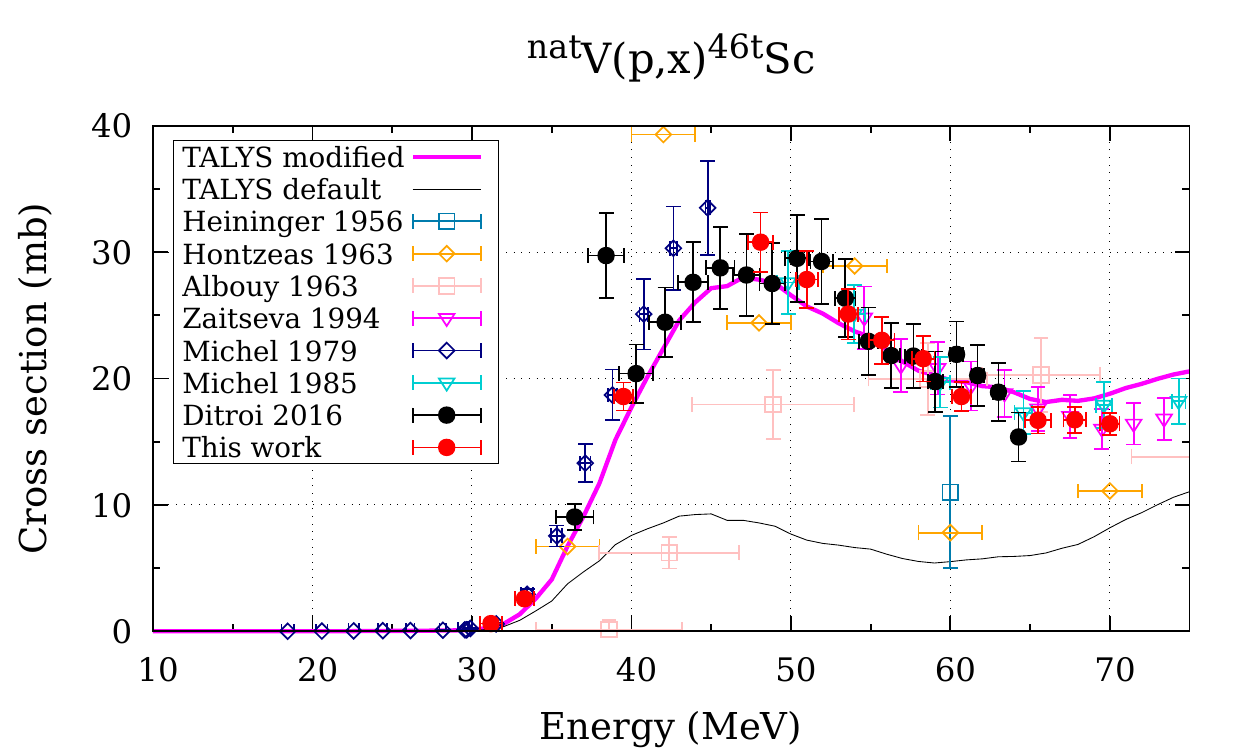}
\caption{\label{fig:sc46xs}$^{46}$Sc cross section. The $^{46t}$Sc notation indicates the cumulative cross section due to the formation of the metastable (half-life: 18.75 s) and ground state of $^{46}$Sc.}
\end{figure}

\begin{figure}[!htb]
\includegraphics[scale=1.0]{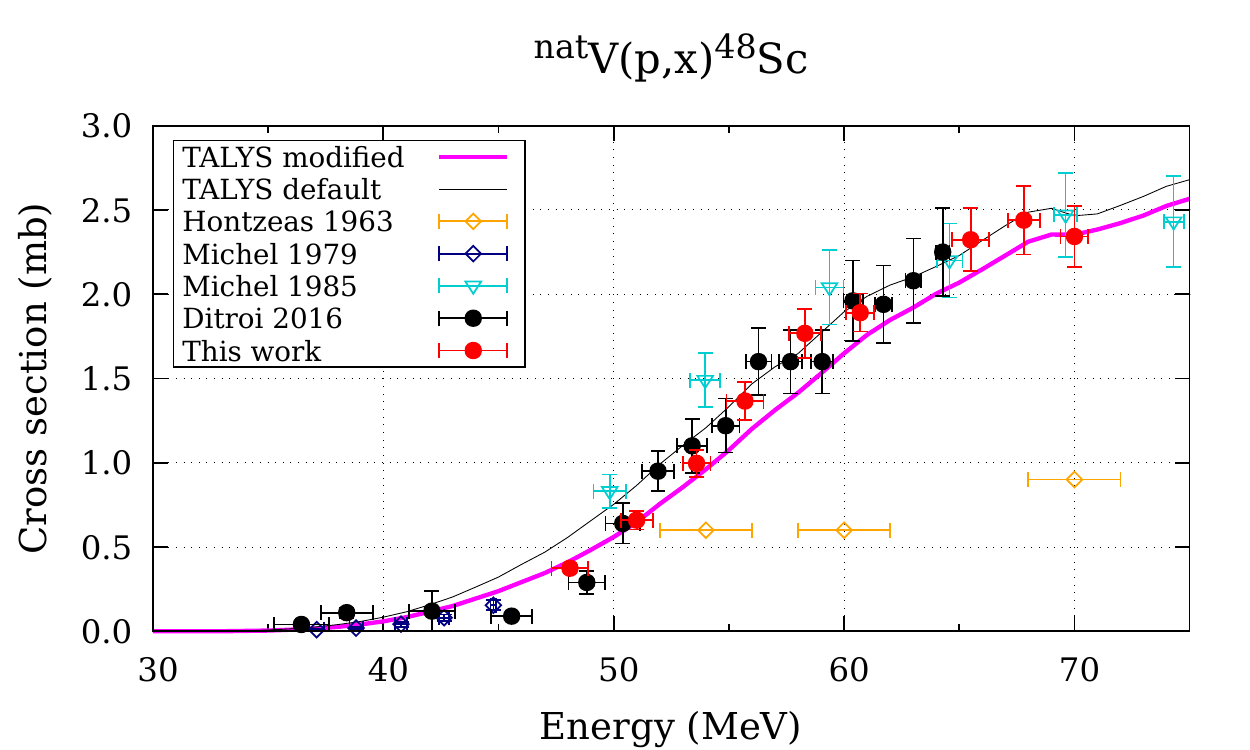}
\caption{\label{fig:sc48xs}$^{48}$Sc cross section.}
\end{figure}

\begin{figure}[!htb]
\includegraphics[scale=1.0]{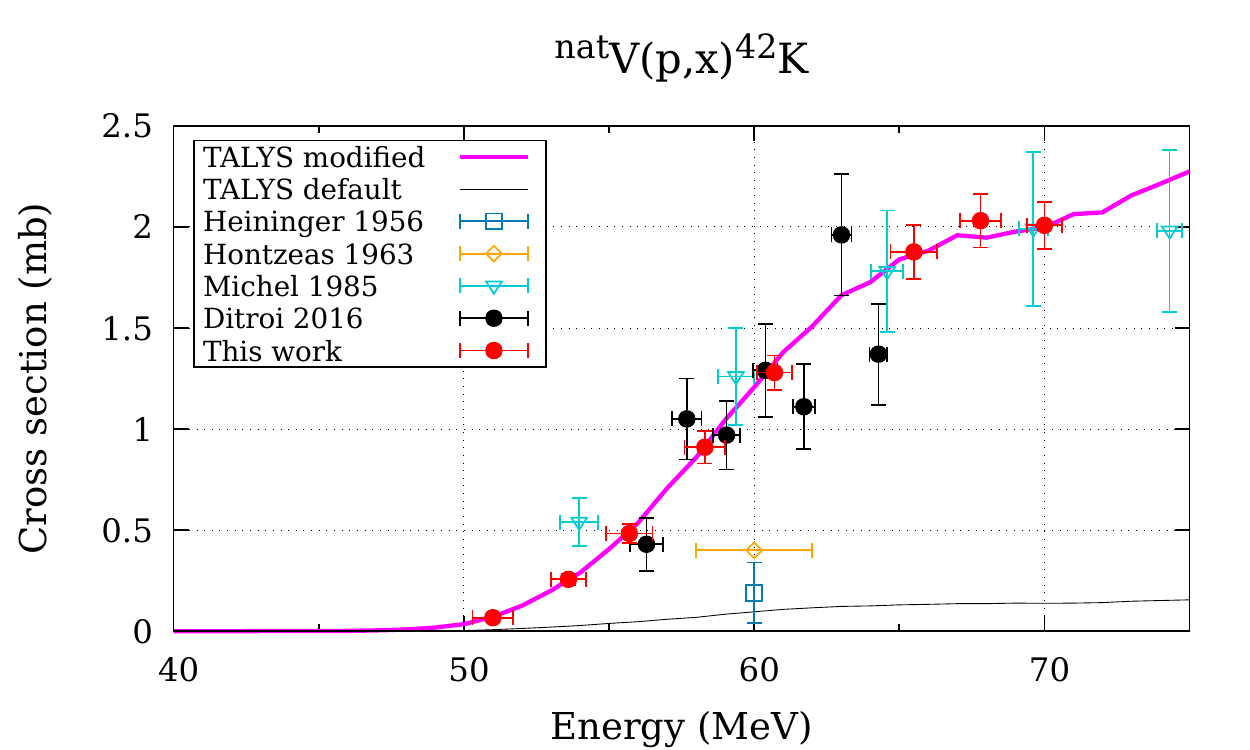}
\caption{\label{fig:k42xs}$^{42}$K cross section.}
\end{figure}

\begin{figure}[!htb]
\includegraphics[scale=1.0]{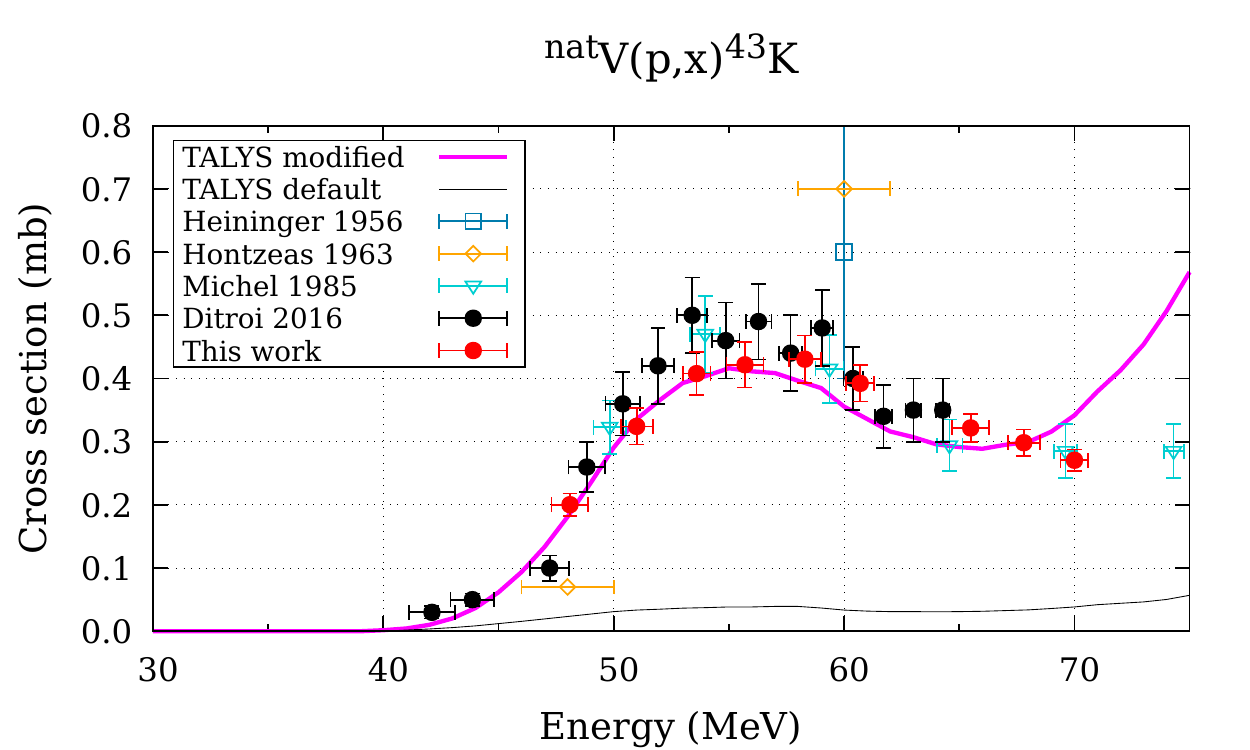}
\caption{\label{fig:k43xs}$^{43}$K cross section.}
\end{figure}

\begin{figure}[!htb]
\includegraphics[scale=1.0]{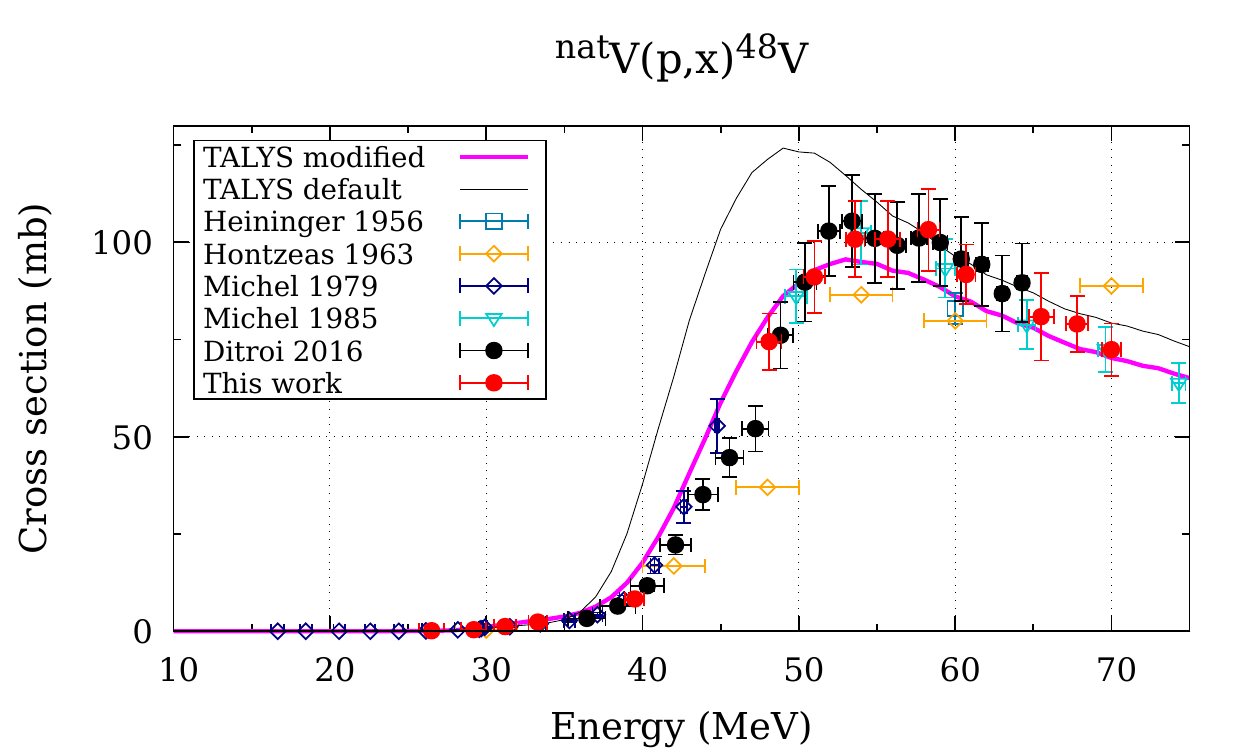}
\caption{\label{fig:v48xs}$^{48}$V cross section.}
\end{figure}

\begin{figure}[!htb]
\includegraphics[scale=1.0]{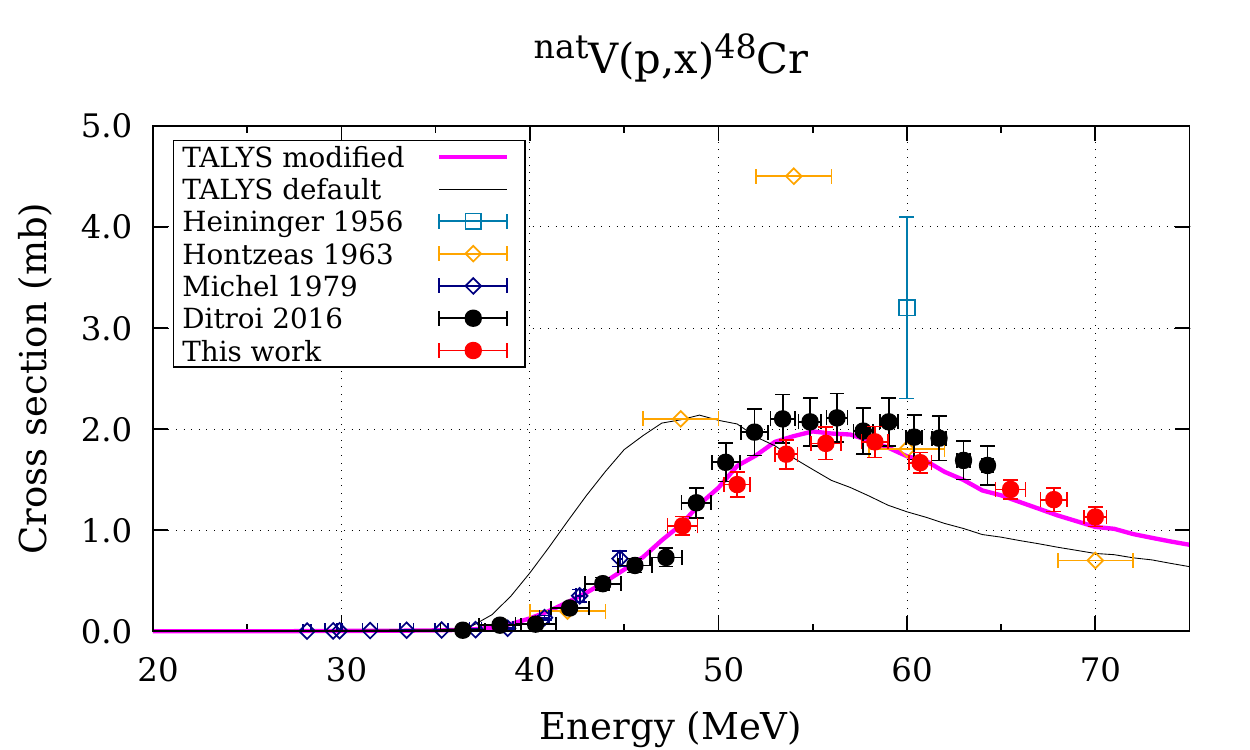}
\caption{\label{fig:cr48xs}$^{48}$Cr cross section.}
\end{figure}

\begin{figure}[!htb]
\includegraphics[scale=1.0]{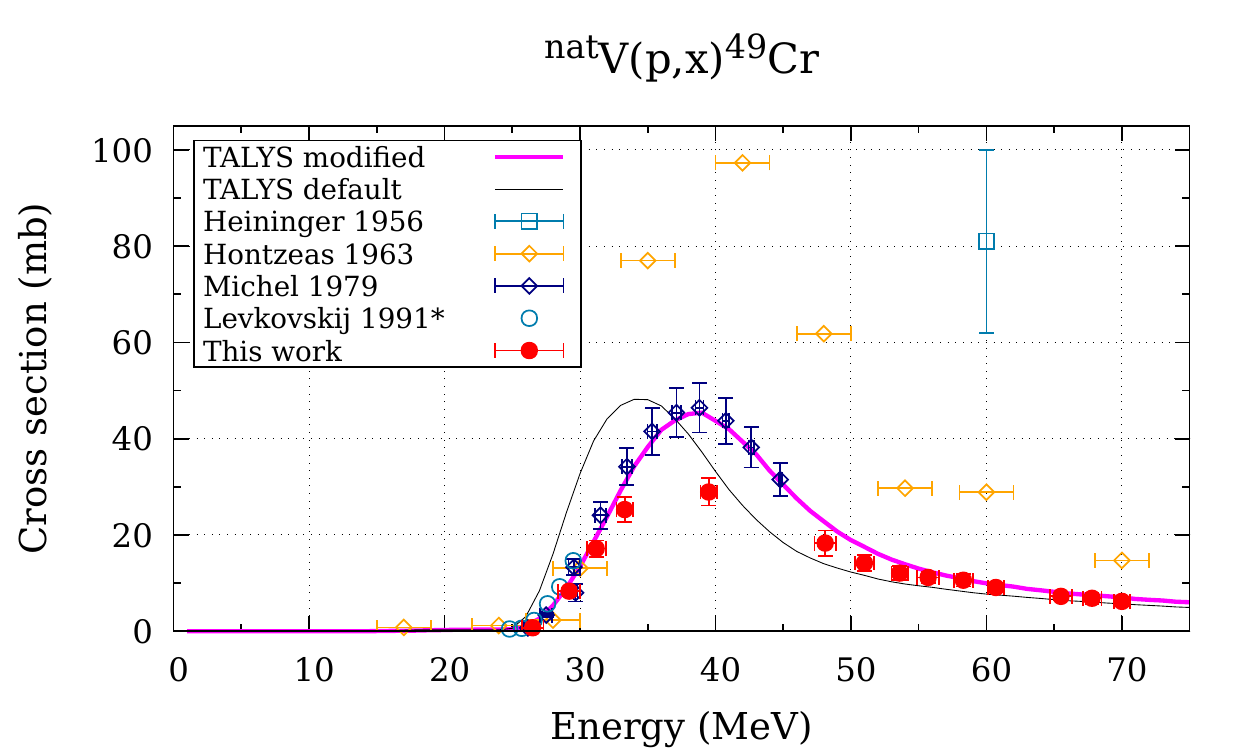}
\caption{\label{fig:cr49xs}$^{49}$Cr cross section.}
\end{figure}

\begin{figure}[!htb]
\includegraphics[scale=1.0]{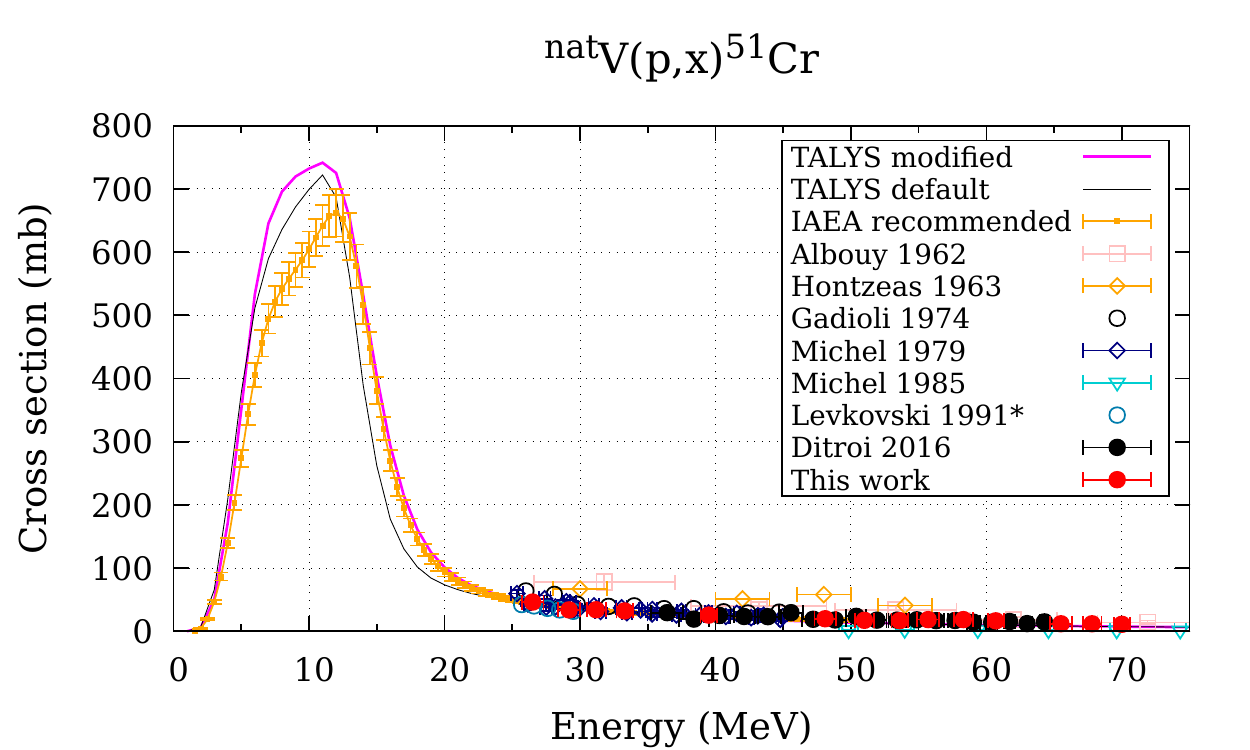}
\caption{\label{fig:cr51xs}$^{51}$Cr cross section. The plot below 25 MeV only shows the IAEA recommended evaluation which represents a guide for the copious number of data measured around the peak.}
\end{figure}

\begin{figure}[!htb]
\includegraphics[scale=1.0]{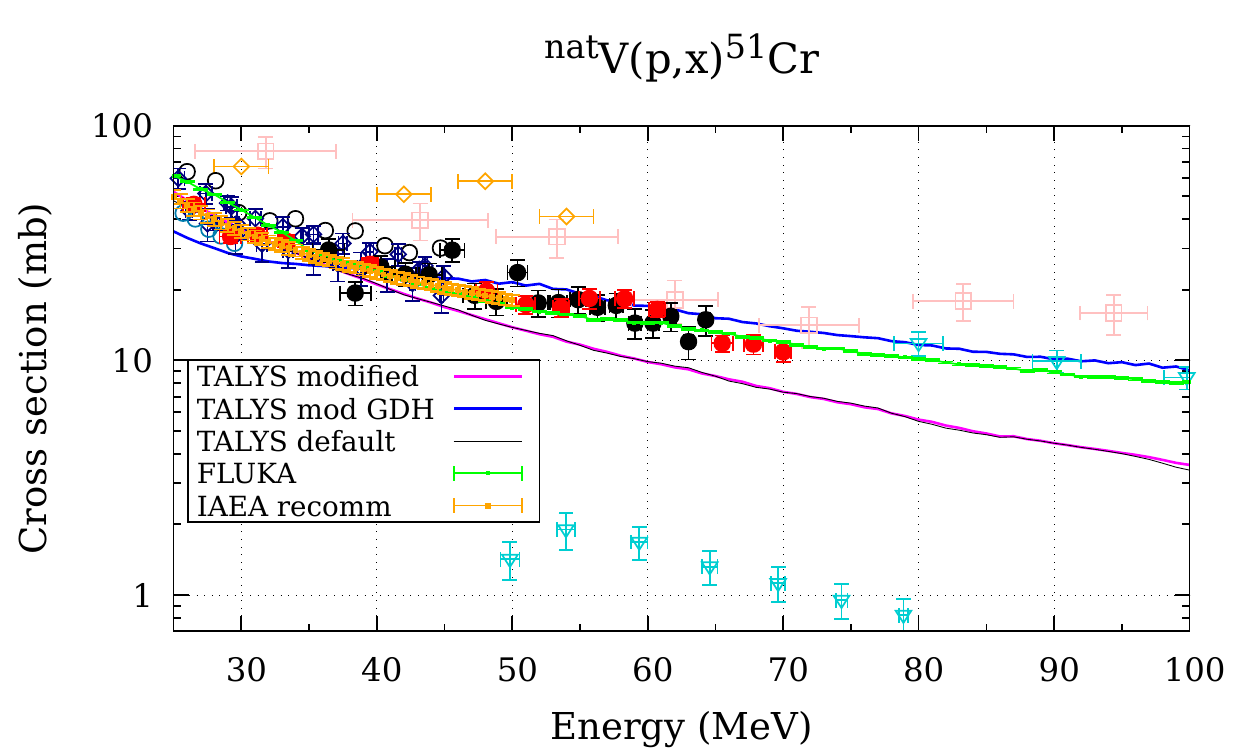}
\caption{\label{fig:cr51xszoom}$^{51}$Cr cross section.}
\end{figure}

This work provides new nuclear reaction data for cross-sections induced by protons on $^{nat}$V targets, within the energy range  26-70 MeV.
Table \ref{tab:3} and Figures \ref{fig:sc47xs} - \ref{fig:cr51xszoom} report the experimental cross section values obtained for the
$^{\mbox{\rm nat}}$V(p,x)$^{47}$Sc, $^{43}$Sc,$^{44g}$Sc, $^{44m}$Sc, $^{46}$Sc, $^{43}$K, $^{42}$K, $^{48}$V, $^{48}$Cr, $^{49}$Cr, $^{51}$Cr reactions. 
The maximum value for the beam energy uncertainty, including the energy straggling, was 0.8 MeV. The
$^{\mbox{\rm nat}}$V(p,x)$^{47}$Sc,$^{46}$Sc
cross section values were already presented in a work dedicated to
$^{47}$Sc production by using
$^{\mbox{\rm nat}}$V targets \cite{pasta2019}. Also the case of
$^{43}$Sc and
$^{43}$K radionuclides was described in a specific
article, since it was found an evident discrepancy with previous results \cite{papersc43}.
The new data are compared with the literature, the default TALYS results {(corresponding to the following options: Koning-Delaroche phenomenological optical potential, exciton model preequilibrium and constant temperature Fermi gas model level density)}, and  the modified TALYS calculation discussed in Section III. The fit turned out remarkably close to the data, not only closer than the default calculation, but also closer than the other predefined models (\textit{ldmodel 4,5,6}), and this has important consequences for the evaluation of the $^{47}$Sc production route for applications in nuclear medicine.

\begin{table}[!hbt]
{\tiny
\begin{tabular}{c|c|c|c|c|c|c|c|c|c|c|c|c}
Energy  &  $^{47}$Sc  &  $^{43}$Sc  &   $^{44g}$Sc  &   $^{44m}$Sc  &  $^{46}$Sc  &  $^{48}$Sc  &   $^{42}$K  &   $^{43}$K  &   $^{48}$V  &  $^{48}$Cr  &   $^{49}$Cr  &   $^{51}$Cr \\ \hline
70.0 $\pm$ 0.6  &  8.0 $\pm$ 0.4   &  0.4 $\pm$ 0.1  &   8.7 $\pm$ 0.6   &  8.0 $\pm$ 0.4  &  16.4 $\pm$ 0.9   &  2.3 $\pm$ 0.2   &  2.0 $\pm$ 0.1  &  0.27 $\pm$ 0.02  &  72.4 $\pm$ 6.7  &  1.1 $\pm$ 0.1   &   6.2 $\pm$ 1.0  &  10.9 $\pm$ 1.0 \\
67.8 $\pm$ 0.7  &  7.7 $\pm$  0.5   &  0.3$\pm$ 0.1  &  7.3$\pm$ 0.6  &  7.0$\pm$ 0.4  &  16.7$\pm$ 1.0   &  2.4 $\pm$ 0.2    &  2.0$\pm$ 0.1  &  0.30$\pm$ 0.02  &  79.0$\pm$ 7.2  & 1.3$\pm$ 0.1   &  6.8$\pm$ 1.1  &  11.7$\pm$ 1.1   \\
65.5 $\pm$ 0.8  &  6.8 $\pm$  0.4   &  0.1$\pm$ 0.1  &  5.7$\pm$ 0.5  &  5.2$\pm$ 0.3  &  16.7$\pm$ 1.0   &  2.3 $\pm$ 0.2     &  1.9$\pm$ 0.1  &  0.32$\pm$ 0.02  &  80.9$\pm$ 11.2  &  1.4$\pm$ 0.1  &  7.2$\pm$ 0.8   &  11.8$\pm$ 0.9 \\
60.7 $\pm$ 0.6  &  5.1 $\pm$  0.3   &  0.3$\pm$ 0.3  &  2.1$\pm$ 0.2  &  1.9$\pm$ 0.1  &  18.6$\pm$ 1.2  &  1.9 $\pm$ 0.1    &  1.3$\pm$ 0.1  &  0.39$\pm$ 0.03  &  91.8$\pm$ 7.6  &  1.7$\pm$ 0.1  &  9.1$\pm$ 1.0   &  16.5$\pm$ 1.2 \\
58.3 $\pm$ 0.7  &  5.0$\pm$ 0.4   &  0.1$\pm$ 0.1  &  1.2$\pm$ 0.1  &  1.1$\pm$ 0.1  &  21.6$\pm$ 1.8  &  1.8 $\pm$ 0.1    &  0.9$\pm$ 0.1  &  0.43$\pm$ 0.04  &  103.3$\pm$ 10.6  &  1.9$\pm$ 0.2  &  10.6$\pm$ 1.3   &  18.3$\pm$ 1.6 \\
55.7 $\pm$ 0.8  &  4.4 $\pm$  0.4  &  0.05$\pm$ 0.03  &  0.6$\pm$ 0.1  &  0.47$\pm$ 0.04  &  23.0$\pm$ 1.9  &  1.4 $\pm$ 0.1  &  0.48$\pm$ 0.05  &  0.42$\pm$ 0.04  &  100.9$\pm$ 9.8  &  1.9$\pm$ 0.2  &  11.2$\pm$ 1.2   &   18.4$\pm$ 1.7\\
53.6 $\pm$ 0.6  &  4.1 $\pm$ 0.3  &  0.07$\pm$ 0.04  &  0.35$\pm$ 0.03   &  0.25$\pm$ 0.02  &  25.1$\pm$ 2.0  &  1.0 $\pm$ 0.1   &  0.26$\pm$ 0.03  &  0.41$\pm$ 0.03  &  100.8$\pm$ 9.8  &  1.8$\pm$ 0.1  &  12.1$\pm$ 1.5  &  16.8$\pm$ 1.4  \\
51.0 $\pm$ 0.7  &  4.1$\pm$ 0.3   &  0.1$\pm$ 0.1  &  0.19$\pm$ 0.04  &  0.13$\pm$ 0.01  &  27.8$\pm$ 2.3   &  0.7 $\pm$ 0.1   &  0.07$\pm$ 0.01  &  0.32$\pm$ 0.03  &  91.1$\pm$ 9.3  &  1.5$\pm$ 0.1   &  14.2$\pm$ 1.7    &  17.3$\pm$ 1.6\\
48.1 $\pm$ 0.8  &  4.6$\pm$ 0.3  &  0.4$\pm$ 0.2  &  0.17$\pm$ 0.05  &  0.05$\pm$ 0.01  & 30.8$\pm$ 2.4   &  0.37 $\pm$ 0.03  &   &  0.20$\pm$ 0.02  & 74.5$\pm$ 7.3   &  1.0$\pm$ 0.1   &  18.3$\pm$ 2.6   &  19.7$\pm$ 1.8 \\
39.5 $\pm$ 0.6  &  8.3$\pm$ 0.5  &    &   &    &   18.6$\pm$ 1.1  &    &    &   &   8.3$\pm$ 0.8  &    &  28.9$\pm$ 2.8   &  25.7$\pm$ 1.6 \\
33.3 $\pm$ 0.6  &  10.9$\pm$ 0.6  &    &    &    &  2.6$\pm$ 0.2   &    &    &    &  2.4$\pm$ 0.2  &    &  25.3$\pm$ 2.6   &  32.3$\pm$ 1.9 \\
31.2 $\pm$ 0.7  &  9.0$\pm$ 0.5  &    &    &    &  0.6$\pm$ 0.1   &    &    &    &  1.2$\pm$ 0.1  &    &  17.2$\pm$ 1.7   &   34.1$\pm$ 2.0\\
29.2 $\pm$ 0.8  &  7.4$\pm$ 0.5  &   &    &    &    &    &    &    &  0.40$\pm$ 0.02   &    &  8.3 $\pm$ 0.9   &  33.7$\pm$ 2.2 \\
26.5 $\pm$ 0.8  &  4.3$\pm$ 0.2   &   &    &    &    &    &    &    &  0.15$\pm$ 0.02  &    &  0.7$\pm$ 0.1   &  46.2$\pm$ 2.9 \\
\end{tabular}
}
\caption{Experimental values of the twelve cross sections measured in this work for the nuclear reactions $^{\mbox{nat}}$V(p,x).}
\label{tab:3}
\end{table}                  

\section{Discussion}

\subsection{On experimental data}

A general good agreement among the new results obtained in this work with the literature data has been found, except with the first data published in 1956 and 1963 \cite{Heininger1956,Hontzeas1963}, probably due to the use of old nuclear and monitor values in those analysis. The largest discrepancy with previous data was found for the $^{43}$Sc case (Fig.\ref{fig:sc43xs}), as described in the specific work that includes also $^{43}$K production (Fig.~\ref{fig:k43xs}) \cite{papersc43}. Since the $^{\mbox{nat}}$V($p,x$)$^{47}$Sc, $^{46t}$Sc cross sections measured within the PASTA project (Figs. \ref{fig:sc47xs} and \ref{fig:sc46xs}) were already presented in \cite{pasta2019}, a detailed discussion is here referred to the production of $^{44}$Sc, $^{44m}$Sc, $^{48}$Sc, $^{48}$V, $^{48}$Cr, $^{49}$Cr, $^{51}$Cr and $^{42}$K. As shown in Figs.~\ref{fig:sc44gxs} and~\ref{fig:sc44mxs}, the cross sections regarding $^{44}$Sc and $^{44m}$Sc are in good agreement with previous data, especially with Ditroi et al.\cite{ditroi2016}. However, our results for $^{44}$Sc at high energies (E $>$ 65 MeV) are slightly higher than the trend indicated by those data up to 60 MeV. The value at 70 MeV obtained by Hontzeas and Yaffe is lower than our result, and the same discrepancy can be found for the $^{44m}$Sc case (Fig.~\ref{fig:sc44mxs}). For the $^{\mbox{nat}}$V($p,x$)$^{44m}$Sc cross section, the results obtained by Michel et al. in 1985~\cite{michel1985} confirm the trend that we have measured. 

In the data analysis of $^{48}$Sc production, only the 1037.52 keV $\gamma$-line was considered, since the 1312.12 keV and 983.526 keV lines are emitted also by $^{48}$V (see Table~\ref{tab:1}). Fig.\ref{fig:sc48xs} reports the new data for the $^{\mbox{nat}}$V($p,x$)$^{48}$Sc cross section, in excellent agreement with the literature \cite{ditroi2016,michel1985} except for data published in 1963~\cite{Hontzeas1963}. 
The $^{\mbox{nat}}$V($p,x$)$^{48}$V cross section, obtained considering the 944.13 keV $\gamma$-line, is shown in Fig.~\ref{fig:v48xs}: also in this case a very good agreement can be noted among the new and previous data in the entire energy range~\cite{michel1979,michel1985,ditroi2016}, also including the oldest values~\cite{Heininger1956,Hontzeas1963}.
		
Fig.~\ref{fig:cr48xs} reports the cross section for $^{48}$Cr production: in the entire energy range an excellent agreement can be seen among our results and the ones published in 2016~\cite{ditroi2016}, that at low energy (E $<$ 45 MeV) are also matching with data by Michel et al.~\cite{michel1979}. Instead, oldest values are quite scattered~\cite{Heininger1956,Hontzeas1963}. 
In the case of $^{49}$Cr (Fig. \ref{fig:cr49xs}) the maximum peak value that we obtained is lower than the one published by R. Michel et al. (1979)~\cite{michel1979}, even if the peak energy seems to correspond. 
Concerning the Levkovski data, as already discussed in \cite{pasta2019}, they
 have been corrected by a factor of 0.77 in Figs.~\ref{fig:sc47xs}, ~\ref{fig:cr49xs} and~\ref{fig:cr51xs} (indicated with a star in the legends). 
The eventual correction for the use of obsolete nuclear data (for the 153 keV $\gamma$-line, they used 42 min half-life and 29.5\% intensity instead of 42.3 min half-life and 30.3\% intensity, see Table \ref{tab:1}) is too small to explain our peak value of about 40\% lower.

The $^{\mbox{nat}}$V($p,x$)$^{51}$Cr cross section measured in this work is in excellent agreement with the extensive literature (Fig.\ref{fig:cr51xszoom}).
Also for the $^{\mbox{nat}}$V($p,x$)$^{42}$K a very good agreement with previous data published by Ditroi et al. and Michel et al. 1985
~\cite{ditroi2016,michel1985}; the previous values at 60 MeV are much lower \cite{Heininger1956,Hontzeas1963}.

\subsection{On theoretical models}

The new calculations for $^{47}$Sc and $^{46}$Sc with the TALYS-modified code (shown in Figs. \ref{fig:sc47xs} and \ref{fig:sc46xs}) greatly improve the agreement with the new measurements, particularly around the peak, where the level density models have an important role. The goodness of the data reproduction is essential in both cases.

In Figs.~\ref{fig:sc43xs}, \ref{fig:k42xs}, and \ref{fig:k43xs}, the new calculation provides a larger cross-section than the default one, and for the two potassium productions, $^{42}$K and $^{43}$K, it reproduces the measured data very well. For $^{43}$Sc, both calculations support the current measurements, however the trend of these new data exhibits an oscillatory behavior that cannot be reproduced by the models and that may be ascribed to statistical fluctuations originating from the low counts of the measurements~\cite{papersc43}.

Additional scandium radionuclides with higher energy thresholds are shown in Figs.~\ref{fig:sc44gxs}, \ref{fig:sc44mxs}, and \ref{fig:sc48xs}. In the case of $^{44g}$Sc (ground) and $^{44m}$Sc
(metastable) the new calculation reproduces the data in the entire energy range while the default one underestimates the data in the higher energy region. For $^{48}$Sc, both calculations reproduce the measurements within the experimental uncertainties, but the TALYS modified result describes better the low-energy rise close to threshold and around 45 MeV.

In Figs.~\ref{fig:v48xs}, \ref{fig:cr48xs}, and \ref{fig:cr49xs}, the improvement of the modified TALYS calculation with respect to the default one is also very significant. All the cross sections calculated with the default option show a clear offset in the peak position, corrected in the new calculation. The changes triggered by the new calculation have been analyzed in detail: the correct reproduction is achieved by the combined effects generated by the optical potential (the JLM optical potential), and by the selected NLD model (\textit{ldmodel 4}), further adjusted with the local optimization of the $c$ and $p$ parameters (modified). It is worth noting that the level density parameters have been fitted to the new data obtained with this work, but, in Fig.~\ref{fig:cr49xs}, the resulting cross section better reproduces the peak value measured in Ref.~\cite{michel1979}.

Finally, in Fig.~\ref{fig:cr51xs}, we provide also the results for $^{51}$Cr production. The new calculation
improves the reproduction of data, particularly in the range 15-20 MeV. This is confirmed in the same region also if we compare the two calculations with the IAEA recommended data~\cite{IAEArecommended}. In the peak range and at lower energies, both calculations show slight discrepancies when compared with the IAEA recommended evaluation. For this reaction, it is interesting to consider the range at higher energy, where the new data have been collected and where the pre-equilibrium mechanisms play a more significant role. This region is analyzed in detail in Fig.~\ref{fig:cr51xszoom}, where it is evident that none of the two calculations (TALYS default and modified) describe correctly the data. We have considered additional new calculations with different pre-equilibrium models in order to seek a better description {in this specific channel}. The recent TALYS-1.95 GDH version~\cite{TALYS-gdh}, in which the Geometry Dependent Hybrid model~\cite{blann1972,Blann1983} was added, has been exploited:  
{this model was previously implemented in other codes, such as ALICE \cite{alice}, and has been included in the TALYS package by the authors \cite{TALYS-gdh}.
}
The calculated excitation function with the GDH model better reproduces the experimental data and this supports the good performance of this newly added pre-equilibrium feature in the TALYS framework. Another reaction code, known to give good description at intermediate energy, is FLUKA~\cite{FLUKA1,FLUKA2}. 
In this code the production of residual nuclei is based on the PEANUT (PreEquilibrium Approach to NUclear Thermalisation) model, which couples a classical intranuclear cascade model, supplemented with quantum corrections, with a preequilibrium framework inspired by the GDH approach. In this framework the produced radionuclides emerge directly from the inelastic hadronic interaction models that are based on a microscopic description of the nucleon-nucleon interaction. This approach is not recommended for the calculation of cross section at very-low energies, close to the reaction thresholds, but, generally, provides reliable results at higher energies. As shown in Fig.~\ref{fig:cr51xszoom}, FLUKA is able to reproduce the $^{51}$Cr cross section high-energy tail with an excellent agreement with the new experimental data.

\begin{figure}[!htb]
\includegraphics[scale=1.0]{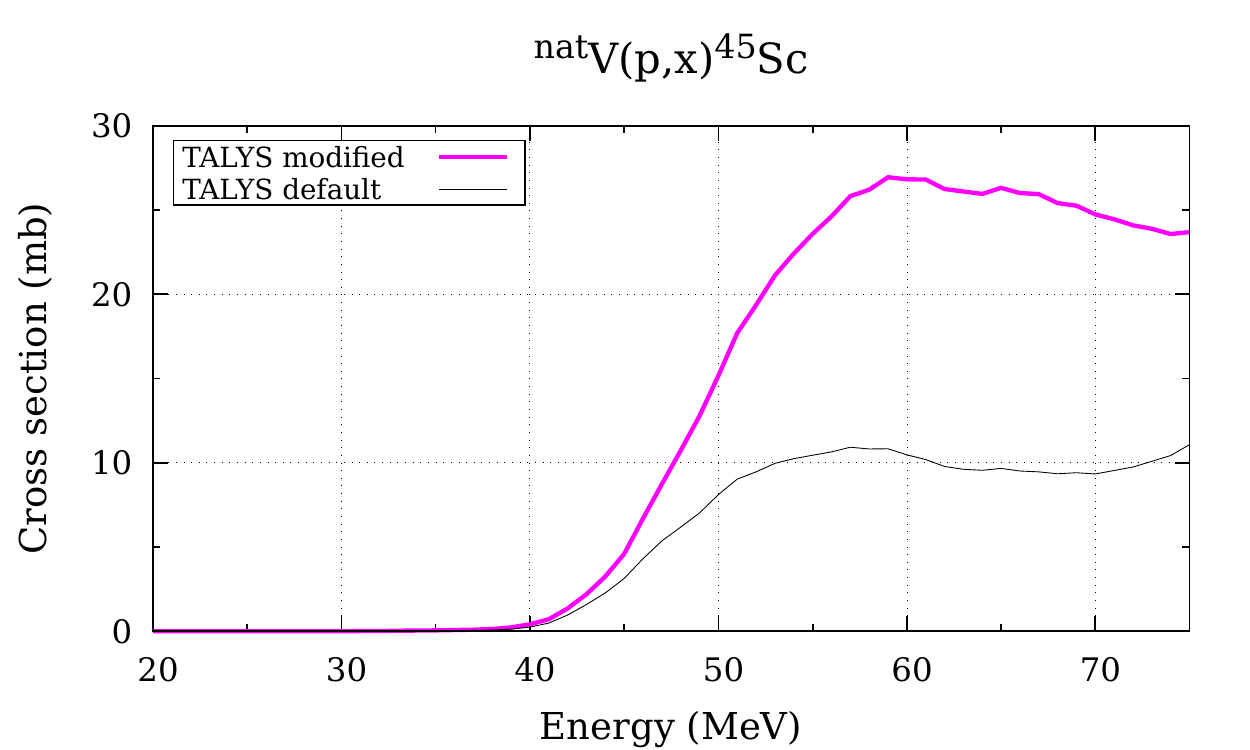}
\caption{\label{fig:sc45xs}$^{45}$Sc cross section.}
\end{figure}

The importance of the theoretical modeling strikingly emerges when one considers the production of stable nuclides, since their cross sections cannot be measured with standard $\gamma$-spectroscopy techniques. Such is the case for the $^{45}${Sc} isotope, shown in Fig.~\ref{fig:sc45xs}. The new cross section exhibits remarkable differences compared to the default one. Without data it is difficult to draw conclusive statements, however, the fact that it provides the correct results for the contiguous reactions (see Figs.~\ref{fig:sc44gxs}, ~\ref{fig:sc44mxs}, ~\ref{fig:sc46xs}) makes the new $^{45}${Sc} result certainly more credible than the TALYS-default result. A reliable theoretical estimate of stable contaminant production is crucial for the production of radio-pharmaceuticals, since $^{45}${Sc} contamination can affect significantly the specific activity of produced $^{47}${Sc}, in the complete absence of relevant cross section data.

\begin{figure}[!htb]
\includegraphics[width=\textwidth]{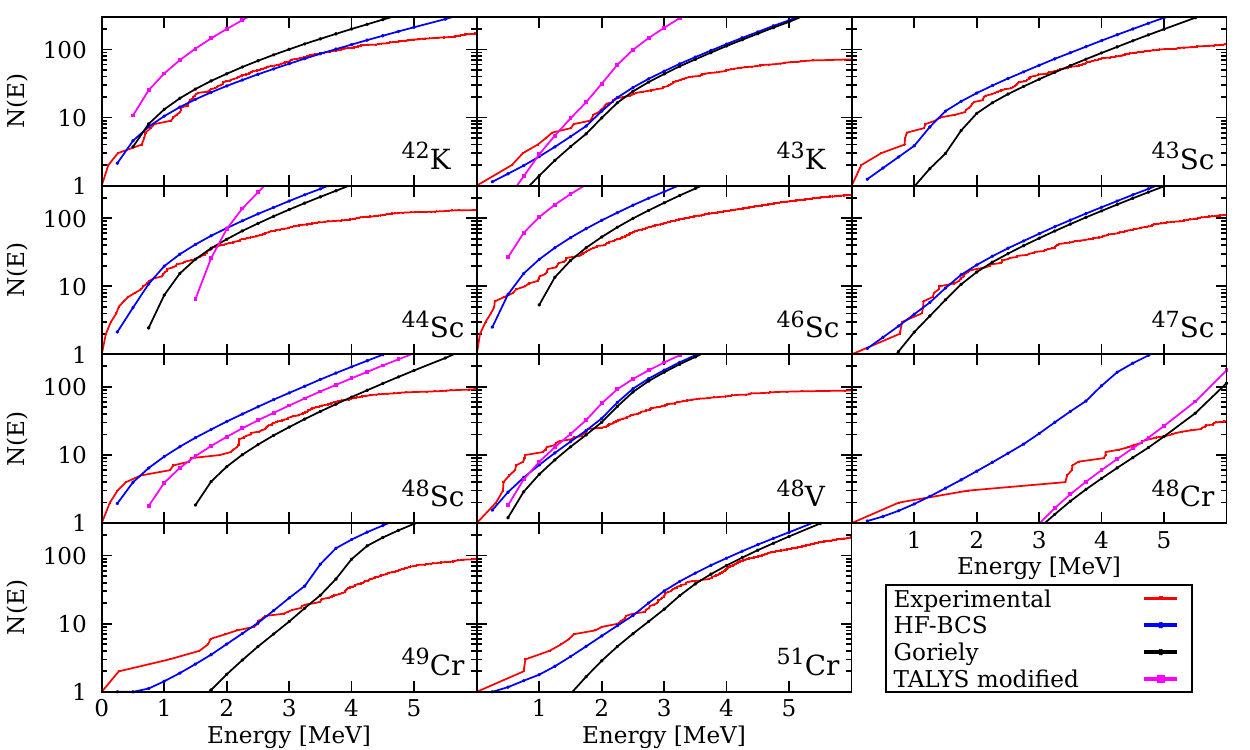}
\caption{Comparison of the observed cumulative of levels with the theoretical models discussed in the text.}
\label{fig:cumul}
\end{figure}

The parameters for the level density found in {the predefined} TALYS {models} are {introduced}  on a global scale, suitable for a broad set of nuclides. In our case we have found a minimal subset of the same parameters that allow to accurately describe the measured channels, without aiming at a self-consistent {global reaction model}. In more detail, we redefined the single $c$ parameter for $^{42}${K} and $^{48}$Cr, the single $p$ parameter for $^{48}$Sc and $^{48}$V and both $c$ and $p$ for  $^{43}$K, $^{44}$Sc and $^{46}$Sc. The corresponding values are given in Tab.~\ref{tab:2}.
This parametrization leads to a solution that optimizes the production of $^{47}$Sc for theranostic applications in agreement with the measurements presented in this work, and allows to accurately calculate the yields and isotopic/radionuclidic purities for realistic irradiation conditions.
Our modified nuclear level density is however different from the previous ones and it is important to assess the effect of the transformation given by Eq.~{\ref{eq:1}} by comparing the theoretical cumulative level distribution with the experimental one deduced from RIPL.  This comparison is presented in Fig.~\ref{fig:cumul}, where the cumulatives from the HF-BCS model are compared with Goriely's global analysis and with the fit obtained in this work.
We stress that the parametrizations given in Tab.~\ref{tab:2} does not improve substantially the agreement between theoretical and experimental cumulatives: that was not our scope. Our aim is to find the best description for the relevant cross sections in view of their importance in radiopharmaceutical applications, and for this reason we have derived the new parameters reported Tab.~\ref{tab:2}.

\subsection{Thick Target Yield (TTY)}

We have calculated the $^{47}$Sc production yield in the same irradiation conditions identified in \cite{denardo2021}. The yield calculations are developed along the lines discussed in Ref.~\cite{canton-fontana-2020}.
{The energy range 19-30 MeV, corresponding to a 1.21 mm thick target perpendicular to the proton beam, can be chosen to avoid $^{46}$Sc co-production. Considering an irradiation time of 24 h and 80 h it is possible to achieve a sufficient amount of $^{47}$Sc with the highest purity to start preclinical and clinical trials.}
The results are shown in Tab.~\ref{tab:4}, where we use the cross section obtained with TALYS default
and with the TALYS modified discussed in this work. 
To complete the  comparison,  we also add the values obtained with the low-energy extrapolation (polynomial fit) discussed in Ref.~\cite{denardo2021} as well as the results obtained with the on-line tool ISOTOPIA, developed at IAEA. We refer to the publications \cite{ISOTOPIA1,ISOTOPIA2} for a detailed description of the simulations provided by ISOTOPIA.

\luciano{
\begin{table}[!htb]
\begin{tabular}{|c|c|c|c|c|c|}
\hline
Radionuclide &   Irradiation time &  TALYS default  & TALYS modified & Polynomial fit & ISOTOPIA \\ 
&  (h) & (MBq/$\mu$A) & (MBq/$\mu$A)  & (MBq/$\mu$A) & (MBq/$\mu$A)\\ \hline
$^{47}$Sc & 24 & 24.43 & 33.48 & 41.50 & 25.29 \\
$^{46}$Sc & 24 & 0.0049  & 0.0137 & 0.0149 & 0.0013\\
$^{47}$Sc & 80 & 65.12  & 89.24 & 111.00 & 67.43 \\
$^{46}$Sc & 80 & 0.0164 & 0.0453 & 0.0492 & 0.0042\\
\hline
\end{tabular}
\caption{Theoretical TTY at EOB of $^{47}Sc$ and its main contaminant $^{46}Sc$ with the different models considered in the analysis for the irradiation profile discussed in the text.}
\label{tab:4}
\end{table}
}

\luciano{
\begin{table}[!htb]
\begin{tabular}{|c|c|c|c|}
\hline
Radionuclide &   Irradiation time &  TALYS default  & TALYS modified  \\ 
&  (h) & (n. of nuclei) & (n. of nuclei) \\ \hline
$^{45}$Sc & 24 & 3.15e9 & 2.07e9    \\
$^{46}$Sc & 24 & 5.18e10 & 1.43e11  \\
$^{47}$Sc & 24 & 1.02e13 & 1.40e13  \\
\hline
\end{tabular}
\caption{Number of nuclei produced after a 24 h irradiation of the thick Vanadium Target. The two radioactive nuclides are compared with the stable isotope $^{45}$Sc.}
\label{tab:5}
\end{table}
}

\begin{figure}[!htb]
\includegraphics[scale=1.0]{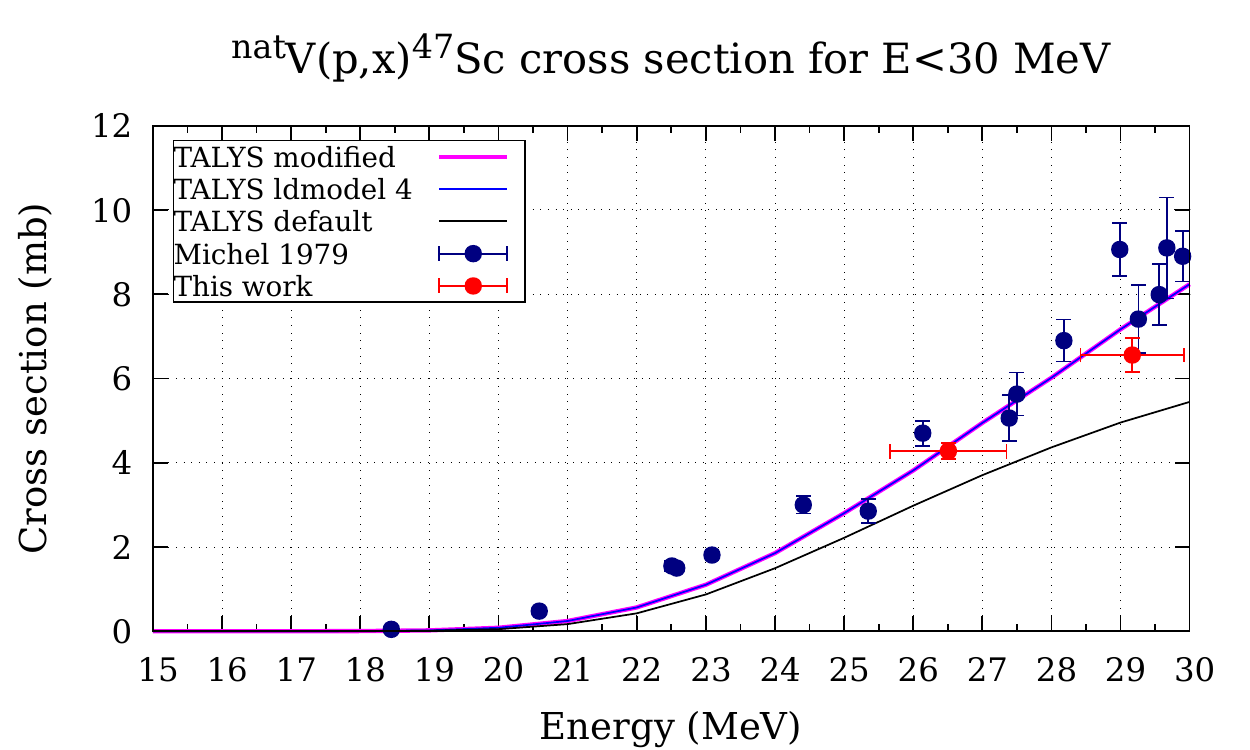}
\caption{\label{fig:sc47zoom}Threshold of the $^{\mbox{\rm nat}}$V(p,x)$^{47}$Sc cross section.}
\end{figure}

\begin{figure}[!htb]
\includegraphics[scale=1.0]{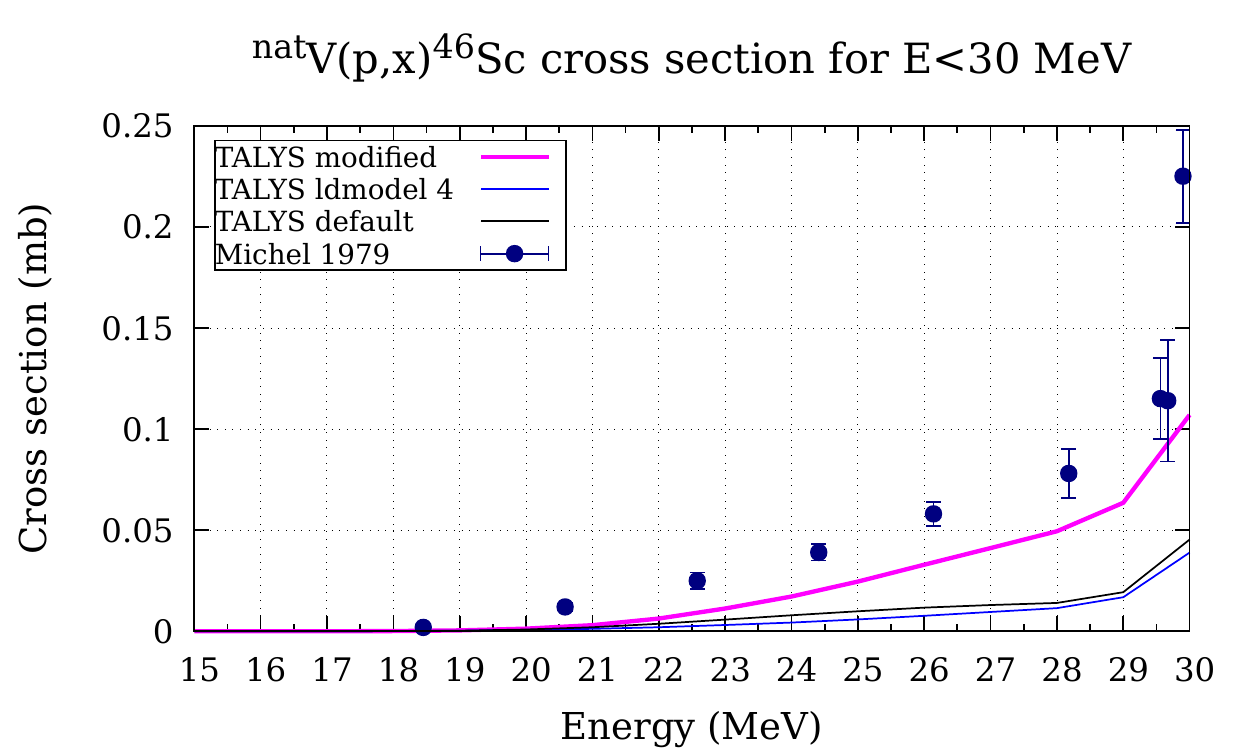}
\caption{\label{fig:sc46zoom}Threshold of the $^{\mbox{\rm nat}}$V(p,x)$^{46}$Sc cross section.}
\end{figure}

From Tab.\ref{tab:4} we observe that, for the $^{47}$Sc yield, TALYS default and ISOTOPIA give very similar results, while for $^{46}$Sc, ISOTOPIA gives estimates four times lower than TALYS default. Both calculations, however, seem to underestimate substantially the $^{47}$Sc yield, and even more severely the yield of the $^{46}$Sc contaminant, when compared with TALYS modified (this work) or with the polynomial fit of Ref.~\cite{denardo2021}. This difference may be ascribed to the fact that both TALYS modified and the polynomial fit adhere more closely to the measurements.
The differences between TALYS modified and the polynomial fit amount to a 21\% increase for the latter, for both $^{47}$Sc and $^{46}$Sc yields. We have inspected the low-energy behaviour of the cross sections in Figs.~\ref{fig:sc47zoom} and \ref{fig:sc46zoom}. For $^{47}$Sc, TALYS modified reproduces the low-energy regime reported by the data in an optimal way, while in the case of $^{46}$Sc, the calculation slightly underestimates the data by Michel~\cite{michel1979}. The low-energy extrapolation of the polynomial fit appears somewhat critical and not completely under control, being built on data sets measured at quite higher energies.
{
On the other hand, the TALYS modified solution is more reliable than the polynomial fit in the threshold region because it contains all the physical effects due to energy (mass threshold) and angular momentum (centifugal barrier) conservation and to the Coulomb barrier: all these constraints are absent in a low energy extrapolation with a polynomial that, for this reason, might be inaccurate. 
}

In Tab.\ref{tab:5} we compare the production (at EOB) of $^{45}$Sc, $^{46}$Sc, and $^{47}$Sc in terms of number of produced nuclei. The impact of $^{45}$Sc contamination appears negligible for both calculations.

\section{Conclusions}
This work reports new data for twelve proton-induced cross sections on $^{\mbox{\rm nat}}$V target that have been measured in the  context of practical applications of novel radiopharmaceutical production.
Along with the experimental measurements, we have studied the phenomenological cross-section modeling to obtain theoretical curves close to the data, as much as possible. The result has been obtained by using TALYS's microscopic models developed in the HF-BCS framework, and by performing a specific optimization of the parameters governing the nuclear levels density. {Concerning the pre-equilibrium models employed, we have selected in TALYS the (default) exciton model with numerically calculated transition rates. This option resulted adequate for the analysis of all cross-sections, with the exception of the high-energy tail of the $^{51}$Cr one. We tested that the GDH pre-equilibrium model, that was recently added to the TALYS package, describes quite well the cross section in this higher energy region, and is comparable with FLUKA results, a nuclear reaction package
suited for modeling cross section in the intermediate energy region.}

This study supports the validity of the approach for nuclear medicine applications and provides an accurate estimation for the yields of $^{47}$Sc and its main contaminants, leading to more reliable assessments of the production route and associated  dosimetric analyses. In spite of some minor differences shown in Tabs.\ref{tab:4}, this work confirms the proton-vanadium route in the 19-30 MeV range as a viable method for producing $^{{47}}$Sc, with a low activity of $^{{46}}$Sc and very minimal contamination of stable $^{{45}}$Sc. It would be interesting, at this point, to confirm these findings with a dedicated TTY experiment with a thick Vanadium target to be irradiated with protons in the suggested energy range.
{The produced $^{{47}}$Sc quantities  are sufficient for pre-clinical and clinical phase trial. To get higher activities, work must be done either on other production routes (either using neutrons or photo-production) or on targetry to accommodate mA proton beams.}

It is important to observe that the tuning of the theoretical level density parameters to obtain accurate cross section is of general applicability for analyzing possible production routes also for other innovative radionuclides for medical applications. 
\begin{acknowledgments}
The authors thanks Dr. Juan Esposito and Dr. Carlos Rossi Alvarez for private communications and the enlightening scientific discussion, the LARAMED team and the ARRONAX staff for the support in the PASTA project. 
This work has been supported by CSN5 of INFN with the PASTA Grant (2017/2018) and partially
supported by EU Horizon 2020 Project RIA-ENSAR2 (654 02) and by
a Grant from the French National Agency for Research called “Investissements
d’Avenir”, Equipex Arronax-Plus no ANR-11-EQPX-0004,
Labex IRON no ANR-11-LABX-18-01 and no ANR-16-IDEX-0007.
\end{acknowledgments}


\bibliography{vnat}

\begin{thebibliography}{54}%
\makeatletter
\providecommand \@ifxundefined [1]{%
 \@ifx{#1\undefined}
}%
\providecommand \@ifnum [1]{%
 \ifnum #1\expandafter \@firstoftwo
 \else \expandafter \@secondoftwo
 \fi
}%
\providecommand \@ifx [1]{%
 \ifx #1\expandafter \@firstoftwo
 \else \expandafter \@secondoftwo
 \fi
}%
\providecommand \natexlab [1]{#1}%
\providecommand \enquote  [1]{``#1''}%
\providecommand \bibnamefont  [1]{#1}%
\providecommand \bibfnamefont [1]{#1}%
\providecommand \citenamefont [1]{#1}%
\providecommand \href@noop [0]{\@secondoftwo}%
\providecommand \href [0]{\begingroup \@sanitize@url \@href}%
\providecommand \@href[1]{\@@startlink{#1}\@@href}%
\providecommand \@@href[1]{\endgroup#1\@@endlink}%
\providecommand \@sanitize@url [0]{\catcode `\\12\catcode `\$12\catcode
  `\&12\catcode `\#12\catcode `\^12\catcode `\_12\catcode `\%12\relax}%
\providecommand \@@startlink[1]{}%
\providecommand \@@endlink[0]{}%
\providecommand \url  [0]{\begingroup\@sanitize@url \@url }%
\providecommand \@url [1]{\endgroup\@href {#1}{\urlprefix }}%
\providecommand \urlprefix  [0]{URL }%
\providecommand \Eprint [0]{\href }%
\providecommand \doibase [0]{https://doi.org/}%
\providecommand \selectlanguage [0]{\@gobble}%
\providecommand \bibinfo  [0]{\@secondoftwo}%
\providecommand \bibfield  [0]{\@secondoftwo}%
\providecommand \translation [1]{[#1]}%
\providecommand \BibitemOpen [0]{}%
\providecommand \bibitemStop [0]{}%
\providecommand \bibitemNoStop [0]{.\EOS\space}%
\providecommand \EOS [0]{\spacefactor3000\relax}%
\providecommand \BibitemShut  [1]{\csname bibitem#1\endcsname}%
\let\auto@bib@innerbib\@empty
\bibitem [{NND()}]{NNDC}%
  \BibitemOpen
  \href@noop {} {\bibinfo {title} {{National Nuclear Data Center (NNDC), NuDat
  2.8 database.}}},\ \bibinfo {howpublished}
  {\url{https://www.nndc.bnl.gov/.}},\ \bibinfo {note} {accessed Mar
  2021}\BibitemShut {NoStop}%
\bibitem [{\citenamefont {Jalilian}\ \emph {et~al.}(2020)\citenamefont
  {Jalilian} \emph {et~al.}}]{jalilian2020iaea}%
  \BibitemOpen
  \bibfield  {author} {\bibinfo {author} {\bibfnamefont {A.~R.}\ \bibnamefont
  {Jalilian}} \emph {et~al.},\ }\bibfield  {title} {\bibinfo {title} {{IAEA
  activities on $^{67}$Cu, $^{186}$Re, $^{47}$Sc Theranostic radionuclides and
  radiopharmaceuticals}},\ }\href@noop {} {\bibfield  {journal} {\bibinfo
  {journal} {Current radiopharmaceuticals}\ }\textbf {\bibinfo {volume} {13}}
  (\bibinfo {year} {2020})}\BibitemShut {NoStop}%
\bibitem [{CRP(2021)}]{CRP}%
  \BibitemOpen
  \href
  {https://www.iaea.org/publications/14793/therapeutic-radiopharmaceuticals-labelled-with-copper-67-rhenium-186-and-scandium-47}
  {\emph {\bibinfo {title} {Therapeutic Radiopharmaceuticals Labelled with
  Copper-67, Rhenium-186 and Scandium-47}}},\ \bibinfo {series} {TECDOC
  Series}\ No.\ \bibinfo {number} {1945}\ (\bibinfo  {publisher} {INTERNATIONAL
  ATOMIC ENERGY AGENCY},\ \bibinfo {address} {Vienna},\ \bibinfo {year}
  {2021})\BibitemShut {NoStop}%
\bibitem [{\citenamefont {M{\"u}ller}\ \emph {et~al.}(2018)\citenamefont
  {M{\"u}ller}, \citenamefont {Domnanich}, \citenamefont {Umbricht},\ and\
  \citenamefont {van~der Meulen}}]{Muller2018}%
  \BibitemOpen
  \bibfield  {author} {\bibinfo {author} {\bibfnamefont {C.}~\bibnamefont
  {M{\"u}ller}}, \bibinfo {author} {\bibfnamefont {K.~A.}\ \bibnamefont
  {Domnanich}}, \bibinfo {author} {\bibfnamefont {C.~A.}\ \bibnamefont
  {Umbricht}},\ and\ \bibinfo {author} {\bibfnamefont {N.~P.}\ \bibnamefont
  {van~der Meulen}},\ }\bibfield  {title} {\bibinfo {title} {Scandium and
  terbium radionuclides for radiotheranostics: current state of development
  towards clinical application},\ }\href@noop {} {\bibfield  {journal}
  {\bibinfo  {journal} {The British journal of radiology}\ }\textbf {\bibinfo
  {volume} {91}},\ \bibinfo {pages} {20180074} (\bibinfo {year}
  {2018})}\BibitemShut {NoStop}%
\bibitem [{\citenamefont {Loveless}\ \emph {et~al.}(2019)\citenamefont
  {Loveless}, \citenamefont {Radford}, \citenamefont {Ferran}, \citenamefont
  {Queern}, \citenamefont {Shepherd},\ and\ \citenamefont
  {Lapi}}]{Loveless2019}%
  \BibitemOpen
  \bibfield  {author} {\bibinfo {author} {\bibfnamefont {C.~S.}\ \bibnamefont
  {Loveless}}, \bibinfo {author} {\bibfnamefont {L.~L.}\ \bibnamefont
  {Radford}}, \bibinfo {author} {\bibfnamefont {S.~J.}\ \bibnamefont {Ferran}},
  \bibinfo {author} {\bibfnamefont {S.~L.}\ \bibnamefont {Queern}}, \bibinfo
  {author} {\bibfnamefont {M.~R.}\ \bibnamefont {Shepherd}},\ and\ \bibinfo
  {author} {\bibfnamefont {S.~E.}\ \bibnamefont {Lapi}},\ }\bibfield  {title}
  {\bibinfo {title} {Photonuclear production, chemistry, and in vitro
  evaluation of the theranostic radionuclide 47 sc},\ }\href@noop {} {\bibfield
   {journal} {\bibinfo  {journal} {EJNMMI research}\ }\textbf {\bibinfo
  {volume} {9}},\ \bibinfo {pages} {1} (\bibinfo {year} {2019})}\BibitemShut
  {NoStop}%
\bibitem [{\citenamefont {Qaim}(2020)}]{Qaim2020}%
  \BibitemOpen
  \bibfield  {author} {\bibinfo {author} {\bibfnamefont {S.~M.}\ \bibnamefont
  {Qaim}},\ }\href@noop {} {\emph {\bibinfo {title} {Medical Radionuclide
  Production: Science and Technology}}}\ (\bibinfo  {publisher} {{De
  Gruyter}},\ \bibinfo {year} {2020})\BibitemShut {NoStop}%
\bibitem [{\citenamefont {Pupillo}\ \emph
  {et~al.}(2019{\natexlab{a}})\citenamefont {Pupillo}, \citenamefont {Mou},
  \citenamefont {Boschi}, \citenamefont {Calzaferri}, \citenamefont {Canton},
  \citenamefont {Cisternino}, \citenamefont {De~Dominicis}, \citenamefont
  {Duatti}, \citenamefont {Fontana}, \citenamefont {Haddad} \emph
  {et~al.}}]{pasta2019}%
  \BibitemOpen
  \bibfield  {author} {\bibinfo {author} {\bibfnamefont {G.}~\bibnamefont
  {Pupillo}}, \bibinfo {author} {\bibfnamefont {L.}~\bibnamefont {Mou}},
  \bibinfo {author} {\bibfnamefont {A.}~\bibnamefont {Boschi}}, \bibinfo
  {author} {\bibfnamefont {S.}~\bibnamefont {Calzaferri}}, \bibinfo {author}
  {\bibfnamefont {L.}~\bibnamefont {Canton}}, \bibinfo {author} {\bibfnamefont
  {S.}~\bibnamefont {Cisternino}}, \bibinfo {author} {\bibfnamefont
  {L.}~\bibnamefont {De~Dominicis}}, \bibinfo {author} {\bibfnamefont
  {A.}~\bibnamefont {Duatti}}, \bibinfo {author} {\bibfnamefont
  {A.}~\bibnamefont {Fontana}}, \bibinfo {author} {\bibfnamefont
  {F.}~\bibnamefont {Haddad}}, \emph {et~al.},\ }\bibfield  {title} {\bibinfo
  {title} {Production of 47sc with natural vanadium targets: results of the
  pasta project},\ }\href@noop {} {\bibfield  {journal} {\bibinfo  {journal}
  {Journal of Radioanalytical and Nuclear Chemistry}\ }\textbf {\bibinfo
  {volume} {322}},\ \bibinfo {pages} {1711} (\bibinfo {year}
  {2019}{\natexlab{a}})}\BibitemShut {NoStop}%
\bibitem [{\citenamefont {Pupillo}\ \emph
  {et~al.}(2019{\natexlab{b}})\citenamefont {Pupillo}, \citenamefont {Fontana},
  \citenamefont {Canton}, \citenamefont {Haddad},\ and\ \citenamefont
  {Skliarova}}]{cimento2018}%
  \BibitemOpen
  \bibfield  {author} {\bibinfo {author} {\bibfnamefont {G.}~\bibnamefont
  {Pupillo}}, \bibinfo {author} {\bibfnamefont {A.}~\bibnamefont {Fontana}},
  \bibinfo {author} {\bibfnamefont {L.}~\bibnamefont {Canton}}, \bibinfo
  {author} {\bibfnamefont {F.}~\bibnamefont {Haddad}},\ and\ \bibinfo {author}
  {\bibfnamefont {H.}~\bibnamefont {Skliarova}},\ }\bibfield  {title} {\bibinfo
  {title} {Preliminary results of the pasta project},\ }\href@noop {}
  {\bibfield  {journal} {\bibinfo  {journal} {Il nuovo cimento C}\ }\textbf
  {\bibinfo {volume} {42}},\ \bibinfo {pages} {1} (\bibinfo {year}
  {2019}{\natexlab{b}})}\BibitemShut {NoStop}%
\bibitem [{\citenamefont {Esposito}\ \emph {et~al.}(2019)\citenamefont
  {Esposito}, \citenamefont {Bettoni}, \citenamefont {Boschi}, \citenamefont
  {Calderolla}, \citenamefont {Cisternino}, \citenamefont {Fiorentini},
  \citenamefont {Keppel}, \citenamefont {Martini}, \citenamefont {Maggiore},
  \citenamefont {Mou}, \citenamefont {Pasquali}, \citenamefont {Pranovi},
  \citenamefont {Pupillo}, \citenamefont {Rossi~Alvarez}, \citenamefont
  {Sarchiapone}, \citenamefont {Sciacca}, \citenamefont {Skliarova},
  \citenamefont {Favaron}, \citenamefont {Lombardi}, \citenamefont {Antonini},\
  and\ \citenamefont {Duatti}}]{laramed2019}%
  \BibitemOpen
  \bibfield  {author} {\bibinfo {author} {\bibfnamefont {J.}~\bibnamefont
  {Esposito}}, \bibinfo {author} {\bibfnamefont {D.}~\bibnamefont {Bettoni}},
  \bibinfo {author} {\bibfnamefont {A.}~\bibnamefont {Boschi}}, \bibinfo
  {author} {\bibfnamefont {M.}~\bibnamefont {Calderolla}}, \bibinfo {author}
  {\bibfnamefont {S.}~\bibnamefont {Cisternino}}, \bibinfo {author}
  {\bibfnamefont {G.}~\bibnamefont {Fiorentini}}, \bibinfo {author}
  {\bibfnamefont {G.}~\bibnamefont {Keppel}}, \bibinfo {author} {\bibfnamefont
  {P.}~\bibnamefont {Martini}}, \bibinfo {author} {\bibfnamefont
  {M.}~\bibnamefont {Maggiore}}, \bibinfo {author} {\bibfnamefont
  {L.}~\bibnamefont {Mou}}, \bibinfo {author} {\bibfnamefont {M.}~\bibnamefont
  {Pasquali}}, \bibinfo {author} {\bibfnamefont {L.}~\bibnamefont {Pranovi}},
  \bibinfo {author} {\bibfnamefont {G.}~\bibnamefont {Pupillo}}, \bibinfo
  {author} {\bibfnamefont {C.}~\bibnamefont {Rossi~Alvarez}}, \bibinfo {author}
  {\bibfnamefont {L.}~\bibnamefont {Sarchiapone}}, \bibinfo {author}
  {\bibfnamefont {G.}~\bibnamefont {Sciacca}}, \bibinfo {author} {\bibfnamefont
  {H.}~\bibnamefont {Skliarova}}, \bibinfo {author} {\bibfnamefont
  {P.}~\bibnamefont {Favaron}}, \bibinfo {author} {\bibfnamefont
  {A.}~\bibnamefont {Lombardi}}, \bibinfo {author} {\bibfnamefont
  {P.}~\bibnamefont {Antonini}},\ and\ \bibinfo {author} {\bibfnamefont
  {A.}~\bibnamefont {Duatti}},\ }\bibfield  {title} {\bibinfo {title} {Laramed:
  A laboratory for radioisotopes of medical interest},\ }\bibfield  {journal}
  {\bibinfo  {journal} {Molecules}\ }\textbf {\bibinfo {volume} {24}},\ \href
  {https://doi.org/10.3390/molecules24010020} {10.3390/molecules24010020}
  (\bibinfo {year} {2019})\BibitemShut {NoStop}%
\bibitem [{\citenamefont {Knapp}\ and\ \citenamefont {Dash}(2016)}]{knapp2016}%
  \BibitemOpen
  \bibfield  {author} {\bibinfo {author} {\bibfnamefont {F.~F.}\ \bibnamefont
  {Knapp}}\ and\ \bibinfo {author} {\bibfnamefont {A.}~\bibnamefont {Dash}},\
  }\href@noop {} {\emph {\bibinfo {title} {Radiopharmaceuticals for therapy}}}\
  (\bibinfo  {publisher} {Springer},\ \bibinfo {year} {2016})\BibitemShut
  {NoStop}%
\bibitem [{IAE(2009)}]{IAEAreport}%
  \BibitemOpen
  \href
  {https://www.iaea.org/publications/7892/cyclotron-produced-radionuclides-physical-characteristics-and-production-methods}
  {\emph {\bibinfo {title} {Cyclotron Produced Radionuclides: Physical
  Characteristics and Production Methods}}},\ \bibinfo {series} {Technical
  Reports Series}\ No.\ \bibinfo {number} {468}\ (\bibinfo  {publisher}
  {INTERNATIONAL ATOMIC ENERGY AGENCY},\ \bibinfo {address} {Vienna},\ \bibinfo
  {year} {2009})\BibitemShut {NoStop}%
\bibitem [{\citenamefont {De~Nardo}\ \emph {et~al.}(2021)\citenamefont
  {De~Nardo}, \citenamefont {Pupillo}, \citenamefont {Mou}, \citenamefont
  {Furlanetto}, \citenamefont {Rosato}, \citenamefont {Esposito},\ and\
  \citenamefont {Mel{\'e}ndez-Alafort}}]{denardo2021}%
  \BibitemOpen
  \bibfield  {author} {\bibinfo {author} {\bibfnamefont {L.}~\bibnamefont
  {De~Nardo}}, \bibinfo {author} {\bibfnamefont {G.}~\bibnamefont {Pupillo}},
  \bibinfo {author} {\bibfnamefont {L.}~\bibnamefont {Mou}}, \bibinfo {author}
  {\bibfnamefont {D.}~\bibnamefont {Furlanetto}}, \bibinfo {author}
  {\bibfnamefont {A.}~\bibnamefont {Rosato}}, \bibinfo {author} {\bibfnamefont
  {J.}~\bibnamefont {Esposito}},\ and\ \bibinfo {author} {\bibfnamefont
  {L.}~\bibnamefont {Mel{\'e}ndez-Alafort}},\ }\bibfield  {title} {\bibinfo
  {title} {Preliminary dosimetric analysis of dota-folate radiopharmaceutical
  radiolabelled with 47sc produced through natv (p, x) 47sc cyclotron
  irradiation},\ }\href@noop {} {\bibfield  {journal} {\bibinfo  {journal}
  {Physics in Medicine \& Biology}\ }\textbf {\bibinfo {volume} {66}},\
  \bibinfo {pages} {025003} (\bibinfo {year} {2021})}\BibitemShut {NoStop}%
\bibitem [{\citenamefont {Heininger}\ and\ \citenamefont
  {Wiig}(1956)}]{Heininger1956}%
  \BibitemOpen
  \bibfield  {author} {\bibinfo {author} {\bibfnamefont {C.~G.}\ \bibnamefont
  {Heininger}}\ and\ \bibinfo {author} {\bibfnamefont {E.~O.}\ \bibnamefont
  {Wiig}},\ }\bibfield  {title} {\bibinfo {title} {Spallation of vanadium with
  60-, 100-, 175-, and 240-mev protons},\ }\href
  {https://doi.org/10.1103/PhysRev.101.1074} {\bibfield  {journal} {\bibinfo
  {journal} {Phys. Rev.}\ }\textbf {\bibinfo {volume} {101}},\ \bibinfo {pages}
  {1074} (\bibinfo {year} {1956})}\BibitemShut {NoStop}%
\bibitem [{\citenamefont {{N. Otuka et al}}(2014)}]{exfor}%
  \BibitemOpen
  \bibfield  {author} {\bibinfo {author} {\bibnamefont {{N. Otuka et al}}},\
  }\bibfield  {title} {\bibinfo {title} {{Towards a more complete and accurate
  experimental nuclear reaction data library ({EXFOR}): international
  collaboration between Nuclear Reaction Data Centres (NRDC)}},\ }\href@noop {}
  {\bibfield  {journal} {\bibinfo  {journal} {Nuclear Data Sheet}\ }\textbf
  {\bibinfo {volume} {120}},\ \bibinfo {pages} {272} (\bibinfo {year}
  {2014})}\BibitemShut {NoStop}%
\bibitem [{\citenamefont {Hontzeas}\ and\ \citenamefont
  {Yaffe}(1963)}]{Hontzeas1963}%
  \BibitemOpen
  \bibfield  {author} {\bibinfo {author} {\bibfnamefont {S.}~\bibnamefont
  {Hontzeas}}\ and\ \bibinfo {author} {\bibfnamefont {L.}~\bibnamefont
  {Yaffe}},\ }\bibfield  {title} {\bibinfo {title} {Interaction of vanadium
  with protons of energies up to 84 mev},\ }\href
  {https://doi.org/10.1139/v63-320} {\bibfield  {journal} {\bibinfo  {journal}
  {Canadian Journal of Chemistry}\ }\textbf {\bibinfo {volume} {41}},\ \bibinfo
  {pages} {2194} (\bibinfo {year} {1963})}\BibitemShut {NoStop}%
\bibitem [{\citenamefont {Michel}\ \emph {et~al.}(1979)\citenamefont {Michel},
  \citenamefont {Brinkmann}, \citenamefont {Weigel},\ and\ \citenamefont
  {Herr}}]{michel1979}%
  \BibitemOpen
  \bibfield  {author} {\bibinfo {author} {\bibfnamefont {R.}~\bibnamefont
  {Michel}}, \bibinfo {author} {\bibfnamefont {G.}~\bibnamefont {Brinkmann}},
  \bibinfo {author} {\bibfnamefont {H.}~\bibnamefont {Weigel}},\ and\ \bibinfo
  {author} {\bibfnamefont {W.}~\bibnamefont {Herr}},\ }\bibfield  {title}
  {\bibinfo {title} {Measurement and hybrid-model analysis of proton-induced
  reactions with v, fe and co},\ }\href
  {https://doi.org/https://doi.org/10.1016/0375-9474(79)90332-4} {\bibfield
  {journal} {\bibinfo  {journal} {Nuclear Physics A}\ }\textbf {\bibinfo
  {volume} {322}},\ \bibinfo {pages} {40} (\bibinfo {year} {1979})}\BibitemShut
  {NoStop}%
\bibitem [{\citenamefont {Michel}\ \emph {et~al.}(1985)\citenamefont {Michel},
  \citenamefont {Peiffer},\ and\ \citenamefont {Stück}}]{michel1985}%
  \BibitemOpen
  \bibfield  {author} {\bibinfo {author} {\bibfnamefont {R.}~\bibnamefont
  {Michel}}, \bibinfo {author} {\bibfnamefont {F.}~\bibnamefont {Peiffer}},\
  and\ \bibinfo {author} {\bibfnamefont {R.}~\bibnamefont {Stück}},\
  }\bibfield  {title} {\bibinfo {title} {Measurement and hybrid model analysis
  of integral excitation functions for proton-induced reactions on vanadium,
  manganese and cobalt up to 200 mev},\ }\href
  {https://doi.org/https://doi.org/10.1016/0375-9474(85)90441-5} {\bibfield
  {journal} {\bibinfo  {journal} {Nuclear Physics A}\ }\textbf {\bibinfo
  {volume} {441}},\ \bibinfo {pages} {617} (\bibinfo {year}
  {1985})}\BibitemShut {NoStop}%
\bibitem [{\citenamefont {Levkovskij}(1991)}]{Levkovskij1991}%
  \BibitemOpen
  \bibfield  {author} {\bibinfo {author} {\bibfnamefont {V.~N.}\ \bibnamefont
  {Levkovskij}},\ }\href@noop {} {\emph {\bibinfo {title} {Cross-Section of
  Medium Mass Nuclide Activation (A = 40–100) by Medium Energy Protons and
  Alpha-Particles (E = 10–50 MeV)}}}\ (\bibinfo  {publisher} {{Inter-Vesi:
  Moscow, Russia}},\ \bibinfo {year} {1991})\BibitemShut {NoStop}%
\bibitem [{\citenamefont {Ditrói}\ \emph {et~al.}(2016)\citenamefont
  {Ditrói}, \citenamefont {Tárkányi}, \citenamefont {Takács},\ and\
  \citenamefont {Hermanne}}]{ditroi2016}%
  \BibitemOpen
  \bibfield  {author} {\bibinfo {author} {\bibfnamefont {F.}~\bibnamefont
  {Ditrói}}, \bibinfo {author} {\bibfnamefont {F.}~\bibnamefont {Tárkányi}},
  \bibinfo {author} {\bibfnamefont {S.}~\bibnamefont {Takács}},\ and\ \bibinfo
  {author} {\bibfnamefont {A.}~\bibnamefont {Hermanne}},\ }\bibfield  {title}
  {\bibinfo {title} {Activation cross-sections of proton induced reactions on
  vanadium in the 37–65mev energy range},\ }\href
  {https://doi.org/https://doi.org/10.1016/j.nimb.2016.05.015} {\bibfield
  {journal} {\bibinfo  {journal} {Nuclear Instruments and Methods in Physics
  Research Section B: Beam Interactions with Materials and Atoms}\ }\textbf
  {\bibinfo {volume} {381}},\ \bibinfo {pages} {16} (\bibinfo {year}
  {2016})}\BibitemShut {NoStop}%
\bibitem [{\citenamefont {{Canton, Luciano}}\ and\ \citenamefont {{Fontana,
  Andrea}}(2020)}]{canton-fontana-2020}%
  \BibitemOpen
  \bibfield  {author} {\bibinfo {author} {\bibnamefont {{Canton, Luciano}}}\
  and\ \bibinfo {author} {\bibnamefont {{Fontana, Andrea}}},\ }\bibfield
  {title} {\bibinfo {title} {Nuclear physics applied to the production of
  innovative radiopharmaceuticals},\ }\href
  {https://doi.org/10.1140/epjp/s13360-020-00730-z} {\bibfield  {journal}
  {\bibinfo  {journal} {Eur. Phys. J. Plus}\ }\textbf {\bibinfo {volume}
  {135}},\ \bibinfo {pages} {770} (\bibinfo {year} {2020})}\BibitemShut
  {NoStop}%
\bibitem [{\citenamefont {Gilbert}\ and\ \citenamefont
  {Cameron}(1965)}]{gilbert-cameron-1965}%
  \BibitemOpen
  \bibfield  {author} {\bibinfo {author} {\bibfnamefont {A.}~\bibnamefont
  {Gilbert}}\ and\ \bibinfo {author} {\bibfnamefont {A.~G.~W.}\ \bibnamefont
  {Cameron}},\ }\bibfield  {title} {\bibinfo {title} {A composite nuclear-level
  density formula with shell corrections},\ }\href
  {https://doi.org/10.1139/p65-139} {\bibfield  {journal} {\bibinfo  {journal}
  {Canadian Journal of Physics}\ }\textbf {\bibinfo {volume} {43}},\ \bibinfo
  {pages} {1446} (\bibinfo {year} {1965})}\BibitemShut {NoStop}%
\bibitem [{\citenamefont {Ignatyuk}\ \emph {et~al.}(1993)\citenamefont
  {Ignatyuk}, \citenamefont {Weil}, \citenamefont {Raman},\ and\ \citenamefont
  {Kahane}}]{ignatyuk-1993}%
  \BibitemOpen
  \bibfield  {author} {\bibinfo {author} {\bibfnamefont {A.~V.}\ \bibnamefont
  {Ignatyuk}}, \bibinfo {author} {\bibfnamefont {J.~L.}\ \bibnamefont {Weil}},
  \bibinfo {author} {\bibfnamefont {S.}~\bibnamefont {Raman}},\ and\ \bibinfo
  {author} {\bibfnamefont {S.}~\bibnamefont {Kahane}},\ }\bibfield  {title}
  {\bibinfo {title} {Density of discrete levels in $^{116}\mathrm{Sn}$},\
  }\href {https://doi.org/10.1103/PhysRevC.47.1504} {\bibfield  {journal}
  {\bibinfo  {journal} {Phys. Rev. C}\ }\textbf {\bibinfo {volume} {47}},\
  \bibinfo {pages} {1504} (\bibinfo {year} {1993})}\BibitemShut {NoStop}%
\bibitem [{\citenamefont {Demetriou}\ and\ \citenamefont
  {Goriely}(2001)}]{demetriou-goriely-2001}%
  \BibitemOpen
  \bibfield  {author} {\bibinfo {author} {\bibfnamefont {P.}~\bibnamefont
  {Demetriou}}\ and\ \bibinfo {author} {\bibfnamefont {S.}~\bibnamefont
  {Goriely}},\ }\bibfield  {title} {\bibinfo {title} {Microscopic nuclear level
  densities for practical applications},\ }\href
  {https://doi.org/https://doi.org/10.1016/S0375-9474(01)01095-8} {\bibfield
  {journal} {\bibinfo  {journal} {Nuclear Physics A}\ }\textbf {\bibinfo
  {volume} {695}},\ \bibinfo {pages} {95} (\bibinfo {year} {2001})}\BibitemShut
  {NoStop}%
\bibitem [{\citenamefont {Minato}(2011)}]{minato-2011}%
  \BibitemOpen
  \bibfield  {author} {\bibinfo {author} {\bibfnamefont {F.}~\bibnamefont
  {Minato}},\ }\bibfield  {title} {\bibinfo {title} {Nuclear level densities
  with microscopic statistical method using a consistent residual
  interaction},\ }\href {https://doi.org/10.1080/18811248.2011.9711785}
  {\bibfield  {journal} {\bibinfo  {journal} {Journal of Nuclear Science and
  Technology}\ }\textbf {\bibinfo {volume} {48}},\ \bibinfo {pages} {984}
  (\bibinfo {year} {2011})}\BibitemShut {NoStop}%
\bibitem [{\citenamefont {Hilaire}\ and\ \citenamefont
  {Goriely}(2006)}]{hilaire2006}%
  \BibitemOpen
  \bibfield  {author} {\bibinfo {author} {\bibfnamefont {S.}~\bibnamefont
  {Hilaire}}\ and\ \bibinfo {author} {\bibfnamefont {S.}~\bibnamefont
  {Goriely}},\ }\bibfield  {title} {\bibinfo {title} {Global microscopic
  nuclear level densities within the hfb plus combinatorial method for
  practical applications},\ }\href
  {https://doi.org/https://doi.org/10.1016/j.nuclphysa.2006.08.014} {\bibfield
  {journal} {\bibinfo  {journal} {Nuclear Physics A}\ }\textbf {\bibinfo
  {volume} {779}},\ \bibinfo {pages} {63} (\bibinfo {year} {2006})}\BibitemShut
  {NoStop}%
\bibitem [{\citenamefont {Goriely}\ \emph
  {et~al.}(2008{\natexlab{a}})\citenamefont {Goriely}, \citenamefont
  {Hilaire},\ and\ \citenamefont {Koning}}]{goriely2008}%
  \BibitemOpen
  \bibfield  {author} {\bibinfo {author} {\bibfnamefont {S.}~\bibnamefont
  {Goriely}}, \bibinfo {author} {\bibfnamefont {S.}~\bibnamefont {Hilaire}},\
  and\ \bibinfo {author} {\bibfnamefont {A.~J.}\ \bibnamefont {Koning}},\
  }\bibfield  {title} {\bibinfo {title} {Improved microscopic nuclear level
  densities within the hartree-fock-bogoliubov plus combinatorial method},\
  }\href@noop {} {\bibfield  {journal} {\bibinfo  {journal} {Physical Review
  C}\ }\textbf {\bibinfo {volume} {78}},\ \bibinfo {pages} {064307} (\bibinfo
  {year} {2008}{\natexlab{a}})}\BibitemShut {NoStop}%
\bibitem [{\citenamefont {Hillman}\ and\ \citenamefont
  {Grover}(1969)}]{hillman-grover-1969}%
  \BibitemOpen
  \bibfield  {author} {\bibinfo {author} {\bibfnamefont {M.}~\bibnamefont
  {Hillman}}\ and\ \bibinfo {author} {\bibfnamefont {J.~R.}\ \bibnamefont
  {Grover}},\ }\bibfield  {title} {\bibinfo {title} {Shell-model combinatorial
  calculations of nuclear level densities},\ }\href
  {https://doi.org/10.1103/PhysRev.185.1303} {\bibfield  {journal} {\bibinfo
  {journal} {Phys. Rev.}\ }\textbf {\bibinfo {volume} {185}},\ \bibinfo {pages}
  {1303} (\bibinfo {year} {1969})}\BibitemShut {NoStop}%
\bibitem [{\citenamefont {Hilaire}\ \emph {et~al.}(2012)\citenamefont
  {Hilaire}, \citenamefont {Girod}, \citenamefont {Goriely},\ and\
  \citenamefont {Koning}}]{hilaire2012}%
  \BibitemOpen
  \bibfield  {author} {\bibinfo {author} {\bibfnamefont {S.}~\bibnamefont
  {Hilaire}}, \bibinfo {author} {\bibfnamefont {M.}~\bibnamefont {Girod}},
  \bibinfo {author} {\bibfnamefont {S.}~\bibnamefont {Goriely}},\ and\ \bibinfo
  {author} {\bibfnamefont {A.~J.}\ \bibnamefont {Koning}},\ }\bibfield  {title}
  {\bibinfo {title} {Temperature-dependent combinatorial level densities with
  the d1m gogny force},\ }\href@noop {} {\bibfield  {journal} {\bibinfo
  {journal} {Physical Review C}\ }\textbf {\bibinfo {volume} {86}},\ \bibinfo
  {pages} {064317} (\bibinfo {year} {2012})}\BibitemShut {NoStop}%
\bibitem [{\citenamefont {Pupillo}\ \emph {et~al.}(2020)\citenamefont
  {Pupillo}, \citenamefont {Mou}, \citenamefont {Haddad}, \citenamefont
  {Fontana},\ and\ \citenamefont {Canton}}]{papersc43}%
  \BibitemOpen
  \bibfield  {author} {\bibinfo {author} {\bibfnamefont {G.}~\bibnamefont
  {Pupillo}}, \bibinfo {author} {\bibfnamefont {L.}~\bibnamefont {Mou}},
  \bibinfo {author} {\bibfnamefont {F.}~\bibnamefont {Haddad}}, \bibinfo
  {author} {\bibfnamefont {A.}~\bibnamefont {Fontana}},\ and\ \bibinfo {author}
  {\bibfnamefont {L.}~\bibnamefont {Canton}},\ }\bibfield  {title} {\bibinfo
  {title} {New results on the natv(p,x)43sc cross section: Analysis of the
  discrepancy with previous data},\ }\href
  {https://doi.org/https://doi.org/10.1016/j.nimb.2019.11.032} {\bibfield
  {journal} {\bibinfo  {journal} {Nuclear Instruments and Methods in Physics
  Research Section B: Beam Interactions with Materials and Atoms}\ }\textbf
  {\bibinfo {volume} {464}},\ \bibinfo {pages} {32} (\bibinfo {year}
  {2020})}\BibitemShut {NoStop}%
\bibitem [{\citenamefont {Haddad}\ \emph {et~al.}(2008)\citenamefont {Haddad},
  \citenamefont {Ferrer}, \citenamefont {Guertin}, \citenamefont {Carlier},
  \citenamefont {Michel}, \citenamefont {Barbet},\ and\ \citenamefont
  {Chatal}}]{haddad2008}%
  \BibitemOpen
  \bibfield  {author} {\bibinfo {author} {\bibfnamefont {F.}~\bibnamefont
  {Haddad}}, \bibinfo {author} {\bibfnamefont {L.}~\bibnamefont {Ferrer}},
  \bibinfo {author} {\bibfnamefont {A.}~\bibnamefont {Guertin}}, \bibinfo
  {author} {\bibfnamefont {T.}~\bibnamefont {Carlier}}, \bibinfo {author}
  {\bibfnamefont {N.}~\bibnamefont {Michel}}, \bibinfo {author} {\bibfnamefont
  {J.}~\bibnamefont {Barbet}},\ and\ \bibinfo {author} {\bibfnamefont {J.-F.}\
  \bibnamefont {Chatal}},\ }\bibfield  {title} {\bibinfo {title} {Arronax, a
  high-energy and high-intensity cyclotron for nuclear medicine},\ }\href@noop
  {} {\bibfield  {journal} {\bibinfo  {journal} {European Journal of Nuclear
  Medicine and Molecular Imaging}\ }\textbf {\bibinfo {volume} {35}},\ \bibinfo
  {pages} {1377} (\bibinfo {year} {2008})}\BibitemShut {NoStop}%
\bibitem [{IAE()}]{IAEAmonitor}%
  \BibitemOpen
  \href@noop {} {\bibinfo {title} {{IAEA Monitor Reactions (2017)}}},\ \bibinfo
  {howpublished}
  {\url{https://www-nds.iaea.org/medical/monitor_reactions.html.}},\ \bibinfo
  {note} {accessed Mar 2019.}\BibitemShut {Stop}%
\bibitem [{\citenamefont {Ziegler}\ \emph {et~al.}(2010)\citenamefont
  {Ziegler}, \citenamefont {Ziegler},\ and\ \citenamefont {Biersack}}]{SRIM}%
  \BibitemOpen
  \bibfield  {author} {\bibinfo {author} {\bibfnamefont {J.~F.}\ \bibnamefont
  {Ziegler}}, \bibinfo {author} {\bibfnamefont {M.~D.}\ \bibnamefont
  {Ziegler}},\ and\ \bibinfo {author} {\bibfnamefont {J.~P.}\ \bibnamefont
  {Biersack}},\ }\bibfield  {title} {\bibinfo {title} {Srim--the stopping and
  range of ions in matter (2010)},\ }\href@noop {} {\bibfield  {journal}
  {\bibinfo  {journal} {Nuclear Instruments and Methods in Physics Research
  Section B: Beam Interactions with Materials and Atoms}\ }\textbf {\bibinfo
  {volume} {268}},\ \bibinfo {pages} {1818} (\bibinfo {year}
  {2010})}\BibitemShut {NoStop}%
\bibitem [{\citenamefont {Otuka}\ \emph {et~al.}(2017)\citenamefont {Otuka},
  \citenamefont {Lalremruata}, \citenamefont {Khandaker}, \citenamefont
  {Usman},\ and\ \citenamefont {Punte}}]{otuka2017}%
  \BibitemOpen
  \bibfield  {author} {\bibinfo {author} {\bibfnamefont {N.}~\bibnamefont
  {Otuka}}, \bibinfo {author} {\bibfnamefont {B.}~\bibnamefont {Lalremruata}},
  \bibinfo {author} {\bibfnamefont {M.}~\bibnamefont {Khandaker}}, \bibinfo
  {author} {\bibfnamefont {A.}~\bibnamefont {Usman}},\ and\ \bibinfo {author}
  {\bibfnamefont {L.}~\bibnamefont {Punte}},\ }\bibfield  {title} {\bibinfo
  {title} {Uncertainty propagation in activation cross section measurements},\
  }\href@noop {} {\bibfield  {journal} {\bibinfo  {journal} {Radiation Physics
  and Chemistry}\ }\textbf {\bibinfo {volume} {140}},\ \bibinfo {pages} {502}
  (\bibinfo {year} {2017})}\BibitemShut {NoStop}%
\bibitem [{\citenamefont {Goriely}\ \emph
  {et~al.}(2008{\natexlab{b}})\citenamefont {Goriely}, \citenamefont
  {Hilaire},\ and\ \citenamefont {Koning}}]{TALYS}%
  \BibitemOpen
  \bibfield  {author} {\bibinfo {author} {\bibfnamefont {S.}~\bibnamefont
  {Goriely}}, \bibinfo {author} {\bibfnamefont {S.}~\bibnamefont {Hilaire}},\
  and\ \bibinfo {author} {\bibfnamefont {A.~J.}\ \bibnamefont {Koning}},\
  }\bibfield  {title} {\bibinfo {title} {{Improved predictions of nuclear
  reaction rates with the TALYS reaction code for astrophysical
  applications}},\ }\href@noop {} {\bibfield  {journal} {\bibinfo  {journal}
  {Astronomy and Astrophysics}\ }\textbf {\bibinfo {volume} {487}},\ \bibinfo
  {pages} {767} (\bibinfo {year} {2008}{\natexlab{b}})}\BibitemShut {NoStop}%
\bibitem [{\citenamefont {{Hilaire, S.}}\ \emph {et~al.}(2010)\citenamefont
  {{Hilaire, S.}}, \citenamefont {{Koning, A.J.}},\ and\ \citenamefont
  {{Goriely, S.}}}]{utopia}%
  \BibitemOpen
  \bibfield  {author} {\bibinfo {author} {\bibnamefont {{Hilaire, S.}}},
  \bibinfo {author} {\bibnamefont {{Koning, A.J.}}},\ and\ \bibinfo {author}
  {\bibnamefont {{Goriely, S.}}},\ }\bibfield  {title} {\bibinfo {title}
  {Microscopic cross sections: An utopia?},\ }\href
  {https://doi.org/10.1051/epjconf/20100802004} {\bibfield  {journal} {\bibinfo
   {journal} {EPJ Web of Conferences}\ }\textbf {\bibinfo {volume} {8}},\
  \bibinfo {pages} {02004} (\bibinfo {year} {2010})}\BibitemShut {NoStop}%
\bibitem [{\citenamefont {Bauge}\ \emph {et~al.}(2001)\citenamefont {Bauge},
  \citenamefont {Delaroche},\ and\ \citenamefont {Girod}}]{bauge2001}%
  \BibitemOpen
  \bibfield  {author} {\bibinfo {author} {\bibfnamefont {E.}~\bibnamefont
  {Bauge}}, \bibinfo {author} {\bibfnamefont {J.~P.}\ \bibnamefont
  {Delaroche}},\ and\ \bibinfo {author} {\bibfnamefont {M.}~\bibnamefont
  {Girod}},\ }\bibfield  {title} {\bibinfo {title} {Lane-consistent,
  semimicroscopic nucleon-nucleus optical model},\ }\href
  {https://doi.org/10.1103/PhysRevC.63.024607} {\bibfield  {journal} {\bibinfo
  {journal} {Phys. Rev. C}\ }\textbf {\bibinfo {volume} {63}},\ \bibinfo
  {pages} {024607} (\bibinfo {year} {2001})}\BibitemShut {NoStop}%
\bibitem [{\citenamefont {Bethe}(1936)}]{bethe}%
  \BibitemOpen
  \bibfield  {author} {\bibinfo {author} {\bibfnamefont {H.~A.}\ \bibnamefont
  {Bethe}},\ }\bibfield  {title} {\bibinfo {title} {An attempt to calculate the
  number of energy levels of a heavy nucleus},\ }\href
  {https://doi.org/10.1103/PhysRev.50.332} {\bibfield  {journal} {\bibinfo
  {journal} {Phys. Rev.}\ }\textbf {\bibinfo {volume} {50}},\ \bibinfo {pages}
  {332} (\bibinfo {year} {1936})}\BibitemShut {NoStop}%
\bibitem [{\citenamefont {Zelevinsky}\ and\ \citenamefont
  {Horoi}(2019)}]{zele}%
  \BibitemOpen
  \bibfield  {author} {\bibinfo {author} {\bibfnamefont {V.}~\bibnamefont
  {Zelevinsky}}\ and\ \bibinfo {author} {\bibfnamefont {M.}~\bibnamefont
  {Horoi}},\ }\bibfield  {title} {\bibinfo {title} {Nuclear level density,
  thermalization, chaos, and collectivity},\ }\href
  {https://doi.org/https://doi.org/10.1016/j.ppnp.2018.12.001} {\bibfield
  {journal} {\bibinfo  {journal} {Progress in Particle and Nuclear Physics}\
  }\textbf {\bibinfo {volume} {105}},\ \bibinfo {pages} {180} (\bibinfo {year}
  {2019})}\BibitemShut {NoStop}%
\bibitem [{\citenamefont {Grimes}(2002)}]{grimes}%
  \BibitemOpen
  \bibfield  {author} {\bibinfo {author} {\bibfnamefont {S.~M.}\ \bibnamefont
  {Grimes}},\ }\bibfield  {title} {\bibinfo {title} {Nuclear level densities},\
  }\href {https://doi.org/10.1080/00223131.2002.10875197} {\bibfield  {journal}
  {\bibinfo  {journal} {Journal of Nuclear Science and Technology}\ }\textbf
  {\bibinfo {volume} {39}},\ \bibinfo {pages} {709} (\bibinfo {year} {2002})},\
  \Eprint
  {https://arxiv.org/abs/https://doi.org/10.1080/00223131.2002.10875197}
  {https://doi.org/10.1080/00223131.2002.10875197} \BibitemShut {NoStop}%
\bibitem [{rip(2009)}]{ripl}%
  \BibitemOpen
  \bibfield  {title} {\bibinfo {title} {{RIPL – Reference Input Parameter
  Library for Calculation of Nuclear Reactions and Nuclear Data Evaluations}},\
  }\href {https://doi.org/https://doi.org/10.1016/j.nds.2009.10.004} {\bibfield
   {journal} {\bibinfo  {journal} {Nuclear Data Sheets}\ }\textbf {\bibinfo
  {volume} {110}},\ \bibinfo {pages} {3107} (\bibinfo {year} {2009})},\
  \bibinfo {note} {special Issue on Nuclear Reaction Data}\BibitemShut
  {NoStop}%
\bibitem [{\citenamefont {Jeukenne}\ \emph
  {et~al.}(1977{\natexlab{a}})\citenamefont {Jeukenne}, \citenamefont
  {Lejeune},\ and\ \citenamefont {Mahaux}}]{jeukenne1977a}%
  \BibitemOpen
  \bibfield  {author} {\bibinfo {author} {\bibfnamefont {J.~P.}\ \bibnamefont
  {Jeukenne}}, \bibinfo {author} {\bibfnamefont {A.}~\bibnamefont {Lejeune}},\
  and\ \bibinfo {author} {\bibfnamefont {C.}~\bibnamefont {Mahaux}},\
  }\bibfield  {title} {\bibinfo {title} {Microscopic calculation of the
  symmetry and coulomb components of the complex optical-model potential},\
  }\href {https://doi.org/10.1103/PhysRevC.15.10} {\bibfield  {journal}
  {\bibinfo  {journal} {Phys. Rev. C}\ }\textbf {\bibinfo {volume} {15}},\
  \bibinfo {pages} {10} (\bibinfo {year} {1977}{\natexlab{a}})}\BibitemShut
  {NoStop}%
\bibitem [{\citenamefont {Jeukenne}\ \emph
  {et~al.}(1977{\natexlab{b}})\citenamefont {Jeukenne}, \citenamefont
  {Lejeune},\ and\ \citenamefont {Mahaux}}]{jeukenne1977b}%
  \BibitemOpen
  \bibfield  {author} {\bibinfo {author} {\bibfnamefont {J.~P.}\ \bibnamefont
  {Jeukenne}}, \bibinfo {author} {\bibfnamefont {A.}~\bibnamefont {Lejeune}},\
  and\ \bibinfo {author} {\bibfnamefont {C.}~\bibnamefont {Mahaux}},\
  }\bibfield  {title} {\bibinfo {title} {Optical-model potential in finite
  nuclei from reid's hard core interaction},\ }\href
  {https://doi.org/10.1103/PhysRevC.16.80} {\bibfield  {journal} {\bibinfo
  {journal} {Phys. Rev. C}\ }\textbf {\bibinfo {volume} {16}},\ \bibinfo
  {pages} {80} (\bibinfo {year} {1977}{\natexlab{b}})}\BibitemShut {NoStop}%
\bibitem [{\citenamefont {{A.J. Koning, S. Hilaire, S.
  Goriely}}(2008)}]{Koning2008}%
  \BibitemOpen
  \bibfield  {author} {\bibinfo {author} {\bibnamefont {{A.J. Koning, S.
  Hilaire, S. Goriely}}},\ }\href@noop {} {\bibfield  {journal} {\bibinfo
  {journal} {Nucl. Phys A}\ }\textbf {\bibinfo {volume} {810}},\ \bibinfo
  {pages} {13} (\bibinfo {year} {2008})}\BibitemShut {NoStop}%
\bibitem [{\citenamefont {{A. Koning, S. Hilaire, S. Goriely}}()}]{TALYSman}%
  \BibitemOpen
  \bibfield  {author} {\bibinfo {author} {\bibnamefont {{A. Koning, S. Hilaire,
  S. Goriely}}},\ }\href@noop {} {\bibinfo {title} {{TALYS-1.9 User Manual}}},\
  \bibinfo {howpublished} {\url{http://www.talys.eu}}\BibitemShut {NoStop}%
\bibitem [{\citenamefont {James}\ and\ \citenamefont {Roos}(1975)}]{minuit}%
  \BibitemOpen
  \bibfield  {author} {\bibinfo {author} {\bibfnamefont {F.}~\bibnamefont
  {James}}\ and\ \bibinfo {author} {\bibfnamefont {M.}~\bibnamefont {Roos}},\
  }\bibfield  {title} {\bibinfo {title} {Minuit: a system for function
  minimization and analysis of the parameter errors and corrections},\
  }\href@noop {} {\bibfield  {journal} {\bibinfo  {journal} {Comput. Phys.
  Commun.}\ }\textbf {\bibinfo {volume} {10}},\ \bibinfo {pages} {343}
  (\bibinfo {year} {1975})}\BibitemShut {NoStop}%
\bibitem [{\citenamefont {T{\'a}rk{\'a}nyi}\ \emph {et~al.}(2019)\citenamefont
  {T{\'a}rk{\'a}nyi}, \citenamefont {Ignatyuk}, \citenamefont {Hermanne},
  \citenamefont {Capote}, \citenamefont {Carlson}, \citenamefont {Engle},
  \citenamefont {Kellett}, \citenamefont {Kibedi}, \citenamefont {Kim},
  \citenamefont {Kondev} \emph {et~al.}}]{IAEArecommended}%
  \BibitemOpen
  \bibfield  {author} {\bibinfo {author} {\bibfnamefont {F.}~\bibnamefont
  {T{\'a}rk{\'a}nyi}}, \bibinfo {author} {\bibfnamefont {A.}~\bibnamefont
  {Ignatyuk}}, \bibinfo {author} {\bibfnamefont {A.}~\bibnamefont {Hermanne}},
  \bibinfo {author} {\bibfnamefont {R.}~\bibnamefont {Capote}}, \bibinfo
  {author} {\bibfnamefont {B.}~\bibnamefont {Carlson}}, \bibinfo {author}
  {\bibfnamefont {J.~W.}\ \bibnamefont {Engle}}, \bibinfo {author}
  {\bibfnamefont {M.~A.}\ \bibnamefont {Kellett}}, \bibinfo {author}
  {\bibfnamefont {T.}~\bibnamefont {Kibedi}}, \bibinfo {author} {\bibfnamefont
  {G.}~\bibnamefont {Kim}}, \bibinfo {author} {\bibfnamefont {F.}~\bibnamefont
  {Kondev}}, \emph {et~al.},\ }\bibfield  {title} {\bibinfo {title}
  {Recommended nuclear data for medical radioisotope production: diagnostic
  gamma emitters},\ }\href@noop {} {\bibfield  {journal} {\bibinfo  {journal}
  {Journal of Radioanalytical and Nuclear Chemistry}\ }\textbf {\bibinfo
  {volume} {319}},\ \bibinfo {pages} {487} (\bibinfo {year}
  {2019})}\BibitemShut {NoStop}%
\bibitem [{\citenamefont {{A.Yu. Konobeyev, U. Fischer, P.E. Pereslavtsev, A.J.
  Koning}}()}]{TALYS-gdh}%
  \BibitemOpen
  \bibfield  {author} {\bibinfo {author} {\bibnamefont {{A.Yu. Konobeyev, U.
  Fischer, P.E. Pereslavtsev, A.J. Koning}}},\ }\bibfield  {title} {\bibinfo
  {title} {{Implementation of GDH model in TALYS-1.7 code}},\ }\href@noop {}
  {\bibinfo  {journal} {{Report KIT Scientific Working Papers 45, 2016}}\
  }\BibitemShut {NoStop}%
\bibitem [{\citenamefont {Blann}(1972)}]{blann1972}%
  \BibitemOpen
\bibfield  {journal} {  }\bibfield  {author} {\bibinfo {author} {\bibfnamefont
  {M.}~\bibnamefont {Blann}},\ }\bibfield  {title} {\bibinfo {title}
  {Importance of the nuclear density distribution on pre-equilibrium decay},\
  }\href@noop {} {\bibfield  {journal} {\bibinfo  {journal} {Physical Review
  Letters}\ }\textbf {\bibinfo {volume} {28}},\ \bibinfo {pages} {757}
  (\bibinfo {year} {1972})}\BibitemShut {NoStop}%
\bibitem [{\citenamefont {Blann}\ and\ \citenamefont
  {Vonach}(1983)}]{Blann1983}%
  \BibitemOpen
  \bibfield  {author} {\bibinfo {author} {\bibfnamefont {M.}~\bibnamefont
  {Blann}}\ and\ \bibinfo {author} {\bibfnamefont {H.~K.}\ \bibnamefont
  {Vonach}},\ }\bibfield  {title} {\bibinfo {title} {Global test of modified
  precompound decay models},\ }\href {https://doi.org/10.1103/PhysRevC.28.1475}
  {\bibfield  {journal} {\bibinfo  {journal} {Phys. Rev. C}\ }\textbf {\bibinfo
  {volume} {28}},\ \bibinfo {pages} {1475} (\bibinfo {year}
  {1983})}\BibitemShut {NoStop}%
\bibitem [{\citenamefont {{C.H.M. Broeders, A.Yu. Konobeyev, A.Yu. Korovin,
  V.P. Lunev, M. Blann}}()}]{alice}%
  \BibitemOpen
  \bibfield  {author} {\bibinfo {author} {\bibnamefont {{C.H.M. Broeders, A.Yu.
  Konobeyev, A.Yu. Korovin, V.P. Lunev, M. Blann}}},\ }\bibfield  {title}
  {\bibinfo {title} {{ALICE/ASH - Pre-compound and evaporation model code
  system for calculation of excitation functions, energy and angular
  distributions of emitted particles in nuclear reactions at intermediate
  energies}},\ }\href@noop {} {\bibinfo  {journal} {{Report FZKA 7183 (May,
  2006)}}\ }\BibitemShut {NoStop}%
\bibitem [{\citenamefont {Böhlen}\ \emph {et~al.}(2014)\citenamefont
  {Böhlen}, \citenamefont {Cerutti}, \citenamefont {Chin}, \citenamefont
  {Fassò}, \citenamefont {Ferrari}, \citenamefont {Ortega}, \citenamefont
  {Mairani}, \citenamefont {Sala}, \citenamefont {Smirnov},\ and\ \citenamefont
  {Vlachoudis}}]{FLUKA1}%
  \BibitemOpen
\bibfield  {journal} {  }\bibfield  {author} {\bibinfo {author} {\bibfnamefont
  {T.}~\bibnamefont {Böhlen}}, \bibinfo {author} {\bibfnamefont
  {F.}~\bibnamefont {Cerutti}}, \bibinfo {author} {\bibfnamefont
  {M.}~\bibnamefont {Chin}}, \bibinfo {author} {\bibfnamefont {A.}~\bibnamefont
  {Fassò}}, \bibinfo {author} {\bibfnamefont {A.}~\bibnamefont {Ferrari}},
  \bibinfo {author} {\bibfnamefont {P.}~\bibnamefont {Ortega}}, \bibinfo
  {author} {\bibfnamefont {A.}~\bibnamefont {Mairani}}, \bibinfo {author}
  {\bibfnamefont {P.}~\bibnamefont {Sala}}, \bibinfo {author} {\bibfnamefont
  {G.}~\bibnamefont {Smirnov}},\ and\ \bibinfo {author} {\bibfnamefont
  {V.}~\bibnamefont {Vlachoudis}},\ }\bibfield  {title} {\bibinfo {title} {The
  fluka code: Developments and challenges for high energy and medical
  applications},\ }\href
  {https://doi.org/https://doi.org/10.1016/j.nds.2014.07.049} {\bibfield
  {journal} {\bibinfo  {journal} {Nuclear Data Sheets}\ }\textbf {\bibinfo
  {volume} {120}},\ \bibinfo {pages} {211} (\bibinfo {year}
  {2014})}\BibitemShut {NoStop}%
\bibitem [{\citenamefont {Ferrari}\ \emph {et~al.}(2005)\citenamefont
  {Ferrari}, \citenamefont {Ranft}, \citenamefont {Sala},\ and\ \citenamefont
  {Fass{\`o}}}]{FLUKA2}%
  \BibitemOpen
  \bibfield  {author} {\bibinfo {author} {\bibfnamefont {A.}~\bibnamefont
  {Ferrari}}, \bibinfo {author} {\bibfnamefont {J.}~\bibnamefont {Ranft}},
  \bibinfo {author} {\bibfnamefont {P.~R.}\ \bibnamefont {Sala}},\ and\
  \bibinfo {author} {\bibfnamefont {A.}~\bibnamefont {Fass{\`o}}},\ }\href@noop
  {} {\emph {\bibinfo {title} {FLUKA: A multi-particle transport code (Program
  version 2005)}}},\ \bibinfo {number} {CERN-2005-10}\ (\bibinfo  {publisher}
  {Cern},\ \bibinfo {year} {2005})\BibitemShut {NoStop}%
\bibitem [{ISO()}]{ISOTOPIA1}%
  \BibitemOpen
  \href@noop {} {\bibinfo {title} {{Medical Isotope Browser}}},\ \bibinfo
  {howpublished}
  {\url{https://www-nds.iaea.org/relnsd/isotopia/isotopia.html.}},\ \bibinfo
  {note} {accessed May 2021.}\BibitemShut {Stop}%
\bibitem [{\citenamefont {Engle}\ \emph {et~al.}(2019)\citenamefont {Engle},
  \citenamefont {Ignatyuk}, \citenamefont {Capote}, \citenamefont {Carlson},
  \citenamefont {Hermanne}, \citenamefont {Kellett}, \citenamefont {Kibedi},
  \citenamefont {Kim}, \citenamefont {Kondev}, \citenamefont {Hussain} \emph
  {et~al.}}]{ISOTOPIA2}%
  \BibitemOpen
  \bibfield  {author} {\bibinfo {author} {\bibfnamefont {J.~W.}\ \bibnamefont
  {Engle}}, \bibinfo {author} {\bibfnamefont {A.~V.}\ \bibnamefont {Ignatyuk}},
  \bibinfo {author} {\bibfnamefont {R.}~\bibnamefont {Capote}}, \bibinfo
  {author} {\bibfnamefont {B.}~\bibnamefont {Carlson}}, \bibinfo {author}
  {\bibfnamefont {A.}~\bibnamefont {Hermanne}}, \bibinfo {author}
  {\bibfnamefont {M.~A.}\ \bibnamefont {Kellett}}, \bibinfo {author}
  {\bibfnamefont {T.}~\bibnamefont {Kibedi}}, \bibinfo {author} {\bibfnamefont
  {G.}~\bibnamefont {Kim}}, \bibinfo {author} {\bibfnamefont {F.~G.}\
  \bibnamefont {Kondev}}, \bibinfo {author} {\bibfnamefont {M.}~\bibnamefont
  {Hussain}}, \emph {et~al.},\ }\bibfield  {title} {\bibinfo {title}
  {Recommended nuclear data for the production of selected therapeutic
  radionuclides},\ }\href@noop {} {\bibfield  {journal} {\bibinfo  {journal}
  {Nuclear Data Sheets}\ }\textbf {\bibinfo {volume} {155}},\ \bibinfo {pages}
  {56} (\bibinfo {year} {2019})}\BibitemShut {NoStop}%
\end{thebibliography}%

\end{document}